\documentclass[preprint,11pt,authoryear]{elsarticle}
\usepackage[top=2.5cm,bottom=2.5cm,left=3cm,right=3cm]{geometry}
\usepackage{natbib}
\usepackage{hyperref}
\makeatletter
\renewcommand\hyper@natlinkbreak[2]{#1}
\makeatother
\usepackage{etoolbox}
\usepackage{tcolorbox}
\usepackage{graphicx}
\setcitestyle{aysep={}} 
\usepackage{amsmath}
\usepackage[hang]{footmisc}
\usepackage{lipsum}
\usepackage{mathrsfs}
\usepackage{bbm}
\usepackage{indentfirst} 
\usepackage{enumerate}
\usepackage{microtype}
\usepackage{geometry}
\usepackage{subfigure}
\usepackage{amssymb}
\usepackage[misc]{ifsym}
\usepackage{mathtools}
\usepackage{soul}
\setstcolor{blue}
\usepackage[pagewise]{lineno}
\usepackage{pdflscape}
\usepackage{setspace}
\usepackage[linesnumbered,ruled,vlined]{algorithm2e}
\usepackage{xcolor}
\usepackage{lmodern} 
\usepackage{graphicx,scalerel}
\usepackage{nameref,hyperref}
\usepackage{caption}
\usepackage{makecell, multirow}
\setcellgapes{3pt}
\usepackage{multirow}
\usepackage{adjustbox}
\usepackage{booktabs}
\usepackage{afterpage}
\usepackage[compact]{titlesec}

\newcommand{\tabitem}{~~\llap{\textbullet}~~}

\newcommand{\Gorder}[1]{}

\newenvironment{stretchpars}
 {\par\setlength{\parfillskip}{0pt}}
 {\par}
 
\newcommand\newhat[1]{\hstretch{2}{\hat{\hstretch{.5}{#1\mkern2mu}}}\mkern-1mu}
\newcommand\newbar[1]{\hstretch{2}{\bar{\hstretch{.5}{#1\mkern2mu}}}\mkern-1mu}

\journal{arXiv}
\begin{document}
%\setstcolor{red}
\begin{frontmatter}



\title{Extinction of bistable populations is affected by the shape of their initial spatial distribution}


\author{Yifei Li$^1$, Stuart T. Johnston$^2$,  Pascal R. Buenzli$^1$, Peter van Heijster$^3$, Matthew J. Simpson$^1$}

\address{$^1$School of Mathematical Sciences, Queensland University of  Technology, Brisbane, Australia.\\
$^2$Systems Biology Laboratory, School of Mathematics and Statistics, and Department of Biomedical Engineering, Melbourne School of Engineering, University of Melbourne, Parkville, Victoria, Australia.\\
       $^3$Biometris, Wageningen University and Research, Wageningen, The Netherlands.\\}

\begin{abstract}
The question of whether biological populations survive or are eventually driven to extinction has long been examined using mathematical models. In this work we study population survival or extinction using a stochastic, discrete lattice-based random walk model where individuals undergo movement, birth and death events. The discrete model is defined on a two-dimensional hexagonal lattice with periodic boundary conditions. A key feature of the discrete model is that crowding effects are introduced by specifying two different crowding functions that govern how local agent density influences movement events and birth/death events. The continuum limit description of the discrete model is a nonlinear reaction-diffusion equation, and we focus on crowding functions that lead to linear diffusion and a bistable source term that is often associated with the strong Allee effect. Using both the discrete and continuum modelling tools we explore the complicated relationship between the long-term survival or extinction of the population and the initial spatial arrangement of the population.  In particular, we study different spatial arrangements of initial distributions: (i) a well-mixed initial distribution where the initial density is independent of position in the domain; (ii) a vertical strip initial distribution where the initial density is independent of vertical position in the domain; and, (iii) several forms of two-dimensional initial distributions where the initial population is distributed in regions with different shapes. Our results indicate that the shape of the initial spatial distribution of the population affects extinction of bistable populations. All software required to solve the discrete and continuum models used in this work are available on 
\href{https://github.com/oneflyli/Yifei2020Dimensionality}{GitHub}.
\end{abstract}


%Highlights:
%\item Shock-fronted travelling waves exist in RDEs with forward-backward diffusion
%\item Different embeddings yield waves with different properties in the same singular limit


\end{frontmatter}


\newpage
\section{Introduction}
\label{intro}
The classical logistic growth model is widely adopted in mathematical biology and mathematical ecology \citep{Kot2001,murray2007mathematical,Keshet2005}. In the logistic model, small initial population densities increase over time to approach a maximum carrying-capacity density \citep{Maini2004cells,Maini2004wound}. An implicit assumption in using the logistic growth model is that any population, no matter how small, will always grow and survive. This limitation also applies to models based on the weak Allee effect, which incorporates a reduced per-capita growth rate relative to the logistic model when the density is small \citep{Taylor2015Allee}. To address this limitation, more complicated models have been developed, including models based on the strong Allee effect \citep{allee1932studies,lewis1993allee,stephens1999allee,COURCHAMP1999405,Taylor2015Allee,Courchamp2008,Arroyo2020}. In the strong Allee effect model, initial densities greater than a threshold, called the \textit{Allee threshold}, grow to eventually reach the carrying capacity, whereas initial densities less than the Allee threshold eventually go extinct \citep{allee1932studies,COURCHAMP1999405,Taylor2015Allee,Courchamp2008,Fadai2020}. This kind of population dynamics, also referred to as  \textit{bistable} population dynamics \citep{Kot2001}, is often adopted to model situations where the potential for population extinction is thought to be important \citep{Saltz1995,COURCHAMP1999405,drake2004,Bottger2015,Vortkamp2020}. Bistable population dynamics are often studied using mathematical models that take the form of an ordinary differential equation (ODE). In this case, the eventual extinction or survival of the population is dictated solely by whether the initial density is greater than, or less than, the Allee threshold density. Such ODE models assume that the population is well-mixed, and hence neglect spatial effects.  Spatial effects, such as moving invasion fronts, can be incorporated by considering partial differential equation (PDE) models where the density of individuals depends explicitly upon position and time \citep{lewis1993allee,holmes1994partial,Hastings2005}. A common PDE framework is to consider a reaction-diffusion equation (RDE) with a cubic bistable source term \citep{neufeld2017role,johnston2017co}. 

When spatial effects are taken into consideration, even the logistic model with linear diffusion may not always lead to the survival of populations. For example, for a population on a finite domain with homogeneous Dirichlet boundary conditions, the population will go extinct when reproduction cannot balance the loss through boundaries \citep{Skellam1951,grindrod1996theory}. The size of the domain must exceed a critical value, called the {\textit{critical patch size}}, so that a population persists \citep{holmes1994partial,lutscher2019integrodifference}. Similar results also hold for diffusing bistable populations, where loss through the boundaries is not the only mechanism of interest since the source term can become negative \citep{BRADFORDa,BRADFORDb}. For a population governed by the strong Allee effect, enough individuals must aggregate together so that the population can reproduce and balance the loss due to the death of individuals. This motivates the concept of the {\textit{critical initial area}} (also known as critical aggregation or critical initial radius) which indicates that the initial population can only survive if the initial occupied area and the initial density are sufficiently large \citep{lewis1993allee,soboleva2003qualitative,lewis2016mathematics}. See Table 1 for a brief review of relevant models and known results.

Current RDE models of bistable populations on two-dimensional domains often consider an infinite domain and a radially symmetric initial distribution \citep{lewis1993allee,petrovskii2001some}. In particular, \citet{lewis1993allee} use formal asymptotics to derive expressions for the critical initial area for a radially distributed bistable population with linear diffusion, and their results are valid in the limit that the time scale of reproduction is much faster than the time scale of migration. In contrast, here we develop a mathematical modelling framework for studying bistable population dynamics on two-dimensional domains with periodic boundary conditions. Using this framework we extend the previous results by showing that bistable populations with the same initial area can either lead to survival or extinction depending upon the initial shape of the population distribution.

\begin{stretchpars}
Our modelling framework is based on a two-dimensional stochastic discrete random walk model on a hexagonal lattice \citep{jin2016stochastic,Fadai2020Proc}. The discrete model is an exclusion process, so that each lattice site can be occupied by no more than one agent. Individuals in the model undergo a birth-death process that is modulated by localised crowding effects \citep{jin2016stochastic,johnston2017co}. The continuum limit of the discrete model leads to a two-dimensional RDE with a bistable source term. This framework allows us to explore discrete simulations together with solutions of the RDE. This approach is convenient because the discrete model is more realistic in the sense that it incorporates fluctuations, but this benefit incurs additional computational overhead~\citep{West2016,Macfarlane2018,Chaplain2020}. Moreover, the discrete framework provides additional information such as the age structure and individual trajectories which cannot be easily obtained using a continuum approach. In contrast, the continuum RDE model can be solved numerically very efficiently,
\end{stretchpars}

\begin{landscape}
\thispagestyle{empty}
{
\begin{table}[]
\vspace{-2cm}
\hspace*{-2cm}
%\centering
\makegapedcells
\begin{adjustbox}{max width=1.7\textwidth}
\begin{tabular}{|p{3.5cm}|p{9.5cm}|p{2.8cm}|p{2.5cm}|p{3.5cm}|p{5.5cm}|p{8.5cm}|}
 \hline
 References & Model & Coordinate & Domain & Boundary conditions & Initial conditions & Properties
 \\
 \hline
 \cite{BRADFORDa}
 & 
 \multirow{2}{9.5cm}{
 \phantom{a}\vfil
 \vspace{-0.4cm}
 \[
     \dfrac{\partial C(x,t)}{\partial t}=\dfrac{\partial^2C(x,t)}{\partial x^2}+f(C(x,t)),
 \]
 where $f$ is a general bistable form.}
  &  
  One-dimensional
  \vfil Cartesian & 
  $0\le x< L$ \vfil where $L<\infty$ \vfil or $L\to\infty$
   &
   \tabitem Homogeneous \vfil \quad\  Neumann at $x=0$ \vfil
      \tabitem Homogeneous \vfil\quad\ or inhomogeneous \vfil\quad\ Dirichlet at $x=L$
   & 
   \tabitem No initial conditions for the \vfil\quad\ steady-state solution $C(x)$. \vfil
   \tabitem Perturbed steady-state \vfil\quad\ solutions for stability analysis.
    & 
   \tabitem There exist stable steady-state solutions, which \vfil\quad\ represent population survival, if $L$ and $C(0)$ are \vfil\quad\ greater than the thresholds.
      \\
      \hline
 \cite{BRADFORDb} &
 \[
     \dfrac{\partial C(r,t)}{\partial t}=\dfrac{1}{r}\dfrac{\partial}{\partial r}\left(r\dfrac{\partial C(r,t)}{\partial r}\right)+f(C(r,t)),
 \]
 where $f$ is a general bistable form.
 & 
 Two-dimensional,\vfil radially symmetric
 &
 $0\le r< R$ \vfil where $R< \infty$ \vfil or $R\to\infty$
 & 
 \tabitem Homogeneous \vfil \quad\  Neumann at $r=0$ \vfil
      \tabitem Homogeneous \vfil\quad\ or inhomogeneous \vfil\quad\ Dirichlet at $r=R$
 & \tabitem No initial conditions for the \vfil\quad\ steady-state solution $C(r)$.\vfil
 \tabitem Perturbed steady-state \vfil\quad\ solutions for stability analysis.
 &
   \tabitem There exist stable steady-state solutions, which  \vfil\quad\ represent population survival, if $L$ and $C(0)$ are \vfil\quad\ greater than the thresholds.\vfil
   \tabitem The threshold of $C(0)$ is significantly greater \vfil\quad\ than  it in \cite{BRADFORDa}.
      \\
 \hline
 \multirow{2}{3cm}{
 \phantom{a}\vfil
 \vspace{-0.2cm}
 \cite{lewis1993allee}}
 & 
 \multirow{2}{9.5cm}{
 \phantom{a}\vfil
 \vspace{-0.2cm}
 \[
     \dfrac{\partial C(x,y,t)}{\partial t}=D\nabla^2C(x,y,t)+kC(1-C)(C-A)
 \]
 }
 & \multirow{2}{2.6cm}{Two-dimensional \vfil Cartesian} & \multirow{2}{2.5cm}{$\mathbb{R}^2$} &  \tabitem Homogeneous \vfil\quad\ Neumann
 & \tabitem $C=1$ in a square region and \vfil\quad\ $C=0$ elsewhere. &  \tabitem Numerical simulations indicate that the initial  \vfil\quad\ distribution converges to a travelling wave \vfil\quad\ solution.
 \\
 \cline{5-7}
 & 
 &  &  &  \tabitem Homogeneous \vfil\quad\ Dirichlet
 & \tabitem $C=1$ in a circular region with \vfil\quad\ radius $r^*$, and $C=0$ elsewhere. & 
 \tabitem There exists a threshold $r_{\text{min}}$ determined by \vfil \quad\ $D,k$ and $A$. If $r^*>r_{\text{min}}$, the initial distribution \vfil \quad\ forms a radially expanding wave which leads to \vfil \quad\ population survival; if $r^*<r_{\text{min}}$, the initial \vfil \quad\ distribution forms a radially shrinking wave \vfil \quad\ which leads to population extinction.
 \\
 \hline
 \cite{soboleva2003qualitative} &
 \[
     \dfrac{\partial C(x,y,t)}{\partial t}=D\nabla^2C(x,y,t)+f(C(x,y,t)),
 \]
 where $f$ is cubic and bistable with $f(0)=f(1)=0$.
 & Two-dimensional \vfil Cartesian &  $\mathbb{R}^2$ &
 \tabitem Homogeneous \vfil\quad\ Dirichlet & 
 \tabitem Perturbed radially symmetric \vfil\quad\ unstable steady-state solutions.\vfil 
 \tabitem Perturbed radially asymmetric \vfil\quad\ unstable steady-state solutions. & 
 \tabitem The unstable steady-state solution provides a \vfil\quad\ threshold initial distribution where populations \vfil\quad\ above the threshold distribution will survive and \vfil\quad\ populations below the threshold distribution will \vfil\quad\ go extinct.\vfil
 \tabitem The symmetric one-dimensional threshold \vfil\quad\ distribution has a smaller maximum density \vfil\quad\ relative to the radially symmetric \vfil\quad\ two-dimensional threshold distribution.
 \\
 \hline
 \cite{kot1996dispersal} &
 Integrodifference equation \vfil (discrete time $n$, continuous space $x$): 
 \[
 C^{n+1}(x)=\int_{-\infty}^{\infty}k(x-\hat{x})f[C^n(\hat{x})]\textls{d}\hat{x},
 \]
 where $f=0$ if $0<C^n<C_\text{A}$ and $f=K$ if $C_\text{A}<C^n<K$. 
   & One-dimensional \vfil Cartesian & $-\infty<x<\infty$ & \tabitem Homogeneous \vfil \quad\ Dirichlet & \tabitem $C=B$ for $-l^*<x<l^*$ \vfil\quad\ and $C=0$ elsewhere.  &
   \tabitem The initial distribution will form an expanding \vfil\quad\ travelling  wave front, which leads to population \vfil\quad\ survival, if $B>C_\text{A}$  and if $l^*$ is greater than a \vfil\quad\ threshold.
   \\ 
 \hline
 \multirow{3}{3.5cm}{
 \phantom{a}\vfil
 \vspace{-1.2cm}
 \cite{etienne2002interaction}} & 
 \multirow{3}{9.5cm}{
 \phantom{a}\vfil
 \vspace{-0.4cm}
 Integrodifference equation \vfil(discrete time $n$, continuous  space $x,y$):
 \[
 L^{n+1}(x,y)=g(R)\iint\limits_{\Omega}k(x-\hat{x},y-\hat{y})A^n(\hat{x},\hat{y})\textls{d}\hat{x}\textls{d}\hat{y},
 \]
 where $A^n=0$ if $L^{n}<L_{\text{min}}$ or $L^{n}>L_{\text{max}}$ and $A^n=L^{n}/2$ elsewhere. Here, $L^n$ is the larval population in generation $n$, $A^n$ is the female adult population in generation $n$ and $g(R)$ represents the resource availability.
 }
 & \multirow{3}{2.6cm}{
 \phantom{a}\vfil
 \vspace{-0.8cm}
 Two-dimensional \vfil Cartesian} &\multirow{3}{2.5cm}{
 \phantom{a}\vfil
 \vspace{-1.2cm}
 $[0,L]\times[0,L]$
 }
 & 
 \tabitem Periodic \vfil
 \phantom{\tabitem} \vfil
 \phantom{\tabitem}
 & \multirow{3}{5.8cm}{
 \phantom{a}\vfil
 \vspace{-0.4cm}
 \tabitem $C=B$ in the whole domain.
 \vfil
 \tabitem $C=B$ in different-sized \vfil\quad\ central square regions and \vfil\quad\ $C=0$ elsewhere.
 }
 & 
 \multirow{3}{8cm}{
 \tabitem Numerical simulations indicate that the  \vfil\quad\ initial distribution and density of individuals, \vfil\quad\ resource availability and heterogeneity \vfil\quad\ influence the fate of populations.
 }
 \\
 \cline{5-5}
 &&&&\tabitem Homogeneous \vfil\quad\  Neumann \vfil
 \phantom{\tabitem}&&
 \\
 \cline{5-5}
 &&&&\tabitem Homogeneous \vfil\quad\ Dirichlet\vfil
 \phantom{\tabitem}&&
 \\
 \hline
 Li et al. (2021) \vfil
 (this paper)&
 Discrete model with the continuum limit 
 \[
     \dfrac{\partial C(x,y,t)}{\partial t}=D\nabla^2 C(x,y,t)+kC(1-C)(C-A)
 \]
 & Two-dimensional \vfil Cartesian &$[0,L]\times[0,L]$ & \tabitem Periodic &  
 \tabitem $C=B$ in the whole domain. 
 \vfil
 \tabitem $C=B$ in regions with different  \vfil\quad\ shapes and $C=0$ elsewhere.
 & \tabitem Both discrete and numerical simulations indicate \vfil\quad\ that initial shapes affect the fate of populations. \vfil\quad\ The key feature of these shapes is their \vfil\quad\ dimensionality.\\
 \hline
\end{tabular}
\end{adjustbox}
\captionsetup{singlelinecheck=off, skip=4pt}
    \caption{The comparison of models studying the critical initial area or critical patch size \citep{BRADFORDa,BRADFORDb} and our model. All models include the strong Allee effect, while \cite{etienne2002interaction} further considers a competition mechanism.}
    \label{table1}
    
    {\makebox[\linewidth]{\thepage}}
\end{table}
}
\end{landscape}
%\restoregeometry

\noindent
but the continuum approach is only accurate if the time scale of migration is small compared to the time scale of proliferation \citep{simpson2010cell}. Moreover, the continuum RDE model does not provide any information about the role of stochasticity~\citep{West2016,Macfarlane2018,Chaplain2020}. So, 
to take advantages of both approaches, we use both a stochastic model and the continuum limit description.

\begin{figure}[]
\centering
\includegraphics[width=\textwidth]{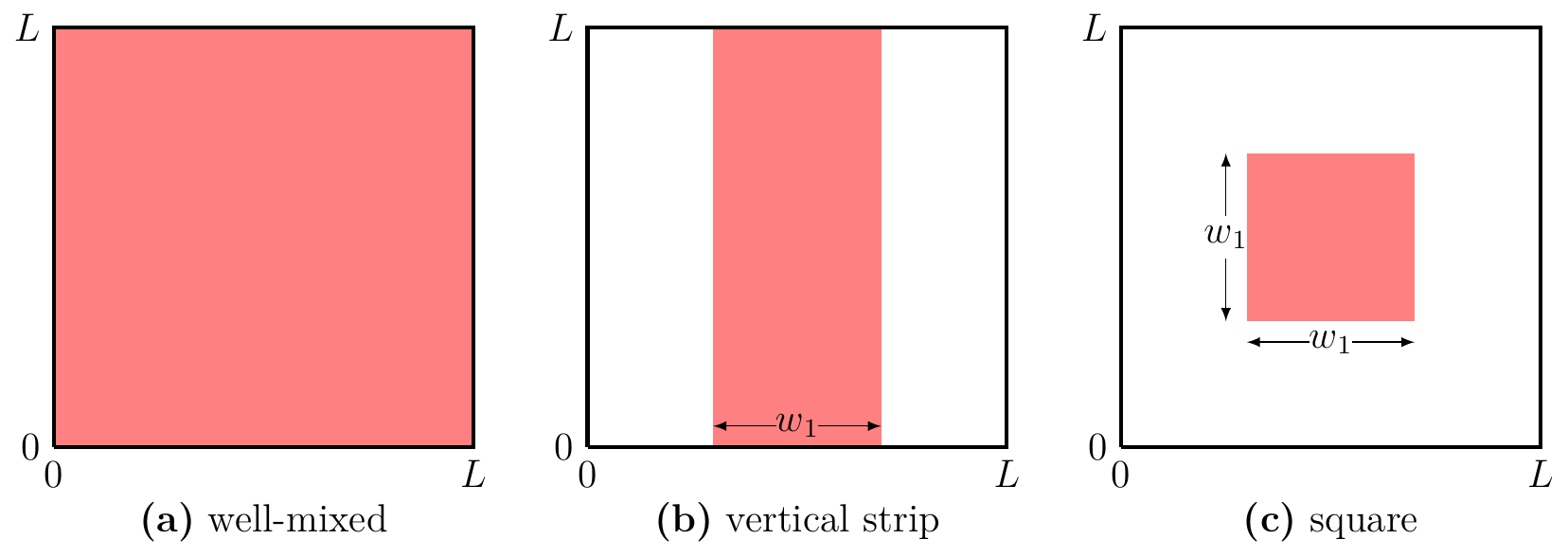}
\singlespace\caption{\textbf{Initial spatial distributions of the population with different shapes on an $L\times L$ square domain.}  In (a), individuals are distributed uniformly across the entire $L\times L$ domain. In (b), individuals are distributed uniformly in a vertical strip of width $w_1$ and height $L$. In (c), individuals are distributed uniformly in the central square region of length and width $w_1$.
}
\label{fig:1} 
\end{figure}

In all cases we study population dynamics on a square domain of side length $L$, with periodic boundary conditions along all boundaries. We explore the role of the initial population distribution by considering different initial spatial arrangements of agents. We first distribute agents uniformly across the entire domain as shown in Figure~{\ref{fig:1}}(a), which leads to a well-mixed population. For the vertical strip initial distribution we distribute agents uniformly within a column of width $w_1$ as shown in Figure~{\ref{fig:1}}(b), which may represent a population of individuals along a one-dimensional river environment \citep{lutscher2010population}. For the initial distributions restricted in both spatial dimensions, we first consider a simple shape and distribute agents uniformly within a square region of area $w_1 \times w_1$ as shown in Figure~{\ref{fig:1}}(c), which may represent a population of cells in a scratch assay \citep{treloar2014vitro}. We further consider several other initial spatial arrangements of agents, see Figures \ref{fig:11}--\ref{fig:13}.

This work is organised as follows. In Section \ref{sec:2} we describe the discrete individual-based model, paying particular attention to incorporating realistic movement and growth mechanisms.  For simplicity, we use the generic term \textit{growth} to refer to the birth/death process in the discrete model. The reason why we make a distinction between birth and death will become clear when we describe the modelling framework. In Section \ref{sec:3} we explain how to analyse the discrete model using a mean-field assumption to arrive at an approximate continuum limit description in terms of a classical RDE. Our discrete-continuous framework incorporates crowding functions into both movement and birth/death mechanisms, which extends the previous work that only considers a crowding function in birth/death mechanisms \citep{jin2016stochastic}. Moreover, our model is very flexible since it describes a wide range of movement and birth/death mechanisms influenced by crowding effects. Results in Section \ref{sec:4} show how both the discrete and continuum models compare. In Section \ref{sec:5}, we systematically explore how population survival or extinction depends upon the shape of the initial distribution. The stochastic lattice random walk model reveals the role of stochasticity in determining the fate of bistable populations. All software required to solve the discrete and continuum models used in this work are available on 
\href{https://github.com/oneflyli/Yifei2020Dimensionality}{GitHub}. 

\section{Discrete model}
\label{sec:2}
We consider a lattice-based discrete model describing movement, birth and death events in a population of individuals on a hexagonal lattice, with lattice spacing $\Delta >0$.  Each lattice site is indexed by $(i,j)$,  and has a unique Cartesian coordinate, 
\begin{equation}
    \label{xyijrelation}
    (x,y)=\left\{\begin{aligned}
    &\left(i\Delta,j\dfrac{\Delta\sqrt{3}}{2}\right),\quad && \text{if $j$ is even,}\\
    &\left(\left(i+\dfrac{1}{2}\right)\Delta,j\dfrac{\Delta\sqrt{3}}{2}\right),\quad && \text{if $j$ is odd.}
    \end{aligned}\right.
\end{equation}
In any single realisation of the stochastic model, a lattice site~$\mathbf{s}$ is either occupied, $C_\mathbf{s}=1$, or vacant, $C_\mathbf{s}=0$. If there are $Q(t)$ agents on the lattice at time $t$, we advance the stochastic simulation from time $t$ to time $t+ \tau$ by randomly selecting $Q(t)$ agents, one at a time, with replacement, so that any particular agent may be selected more than once, and allowing those agents to \textit{attempt} to move.  Once the $Q(t)$ potential movement events have been assessed, we then select  $Q(t)$ agents at random, one at a time, with replacement, to \textit{attempt} to undergo a growth event, which could be either a birth or death event depending upon the local crowding conditions. Although altering the order of these events leads to different outcomes in particular discrete simulations, these differences are not important when we consider averaged data from many identically-prepared realisations of the model \citep{Mat2009MultiExclusion,simpson2009pathlines}.

\begin{figure}
\centering
\includegraphics[width=\textwidth]{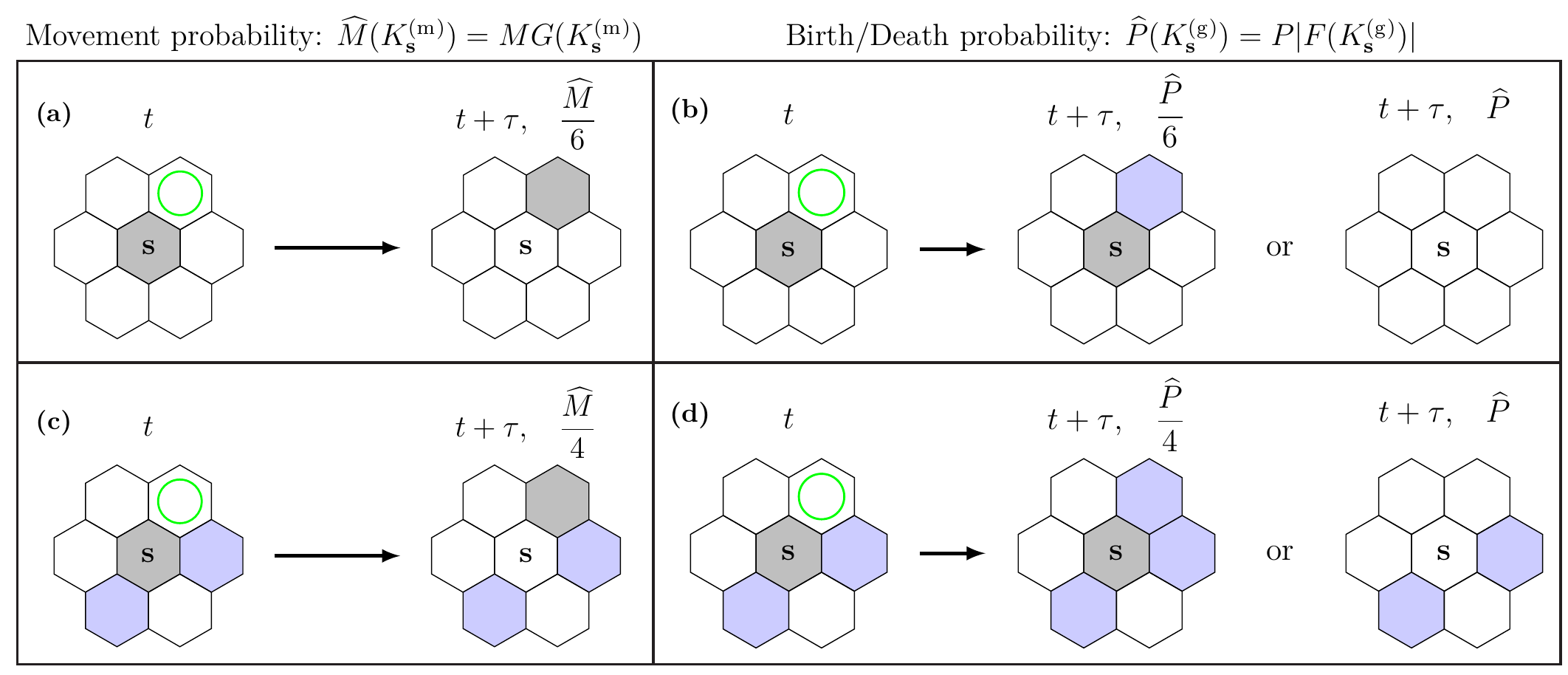}
\singlespace\caption{\textbf{Movement and birth/death mechanisms.} In each lattice fragment site~$\mathbf{s}$ is occupied and shaded in grey, and occupied neighbouring sites are shaded in blue, while vacant neighbouring sites are unshaded (white). In (a) the agent at site~$\mathbf{s}$ moves with probability $\widehat{M}$ and moves to the target site, highlighted with a green circle, with probability $\widehat{M}/6$. In (b) the agent at site~$\mathbf{s}$ undergoes a birth event with probability $\widehat{P}$ and places a new agent on the target site with probability $\widehat{P}/6$ if the growth crowding function $F>0$. In contrast, it dies with probability $\widehat{P}$ if $F<0$. In (c) the agent moves with probability $\widehat{M}$ and moves to the target site with probability $\widehat{M}/4$. In (d) the agent undergoes a birth event with probability $\widehat{P}$ and places a new agent on the target site with probability $\widehat{P}/4$ if $F>0$. In contrast, it dies with probability $\widehat{P}$ if $F<0$. }
\label{fig:2} 
\end{figure}

We now explain some features of the discrete model in terms of the schematic in Figure~\ref{fig:2}. In this initial description of the discrete model we consider nearest-neighbour movement and growth events only, and we will relax this assumption later. Figure~\ref{fig:2}(a) shows a potential movement event for an agent at site $\textbf{s}$, where all nearest-neighbour sites are vacant. In this case, the probability of attempting to move during the next time step of duration $\tau$, is $M \in [0,1]$, and the attempted motility event will be successful with probability $\newhat{M}\le M$. Here we note that the two probabilities, $M$ and $\newhat{M}$ are, in general, different.  This difference is a result of the local crowding effects.  The special case in Figure~\ref{fig:2}(a) where the agent at site $\textbf{s}$ is uncrowded we have $\newhat{M} =  M$. If the attempted motility event is successful, the agent at site $s$  moves to a randomly-chosen vacant site chosen among the set of vacant nearest-neighbour sites.  In this case, as all six neighbour sites are vacant, the probability of moving to the target site, highlighted with a green circle, is $\newhat{M}/6$. 

In Figure~\ref{fig:2}(b) we show a potential growth event for an  agent at site $\textbf{s}$, where again all nearest-neighbour sites are vacant. Here, the probability of attempting to grow in the next time step of duration $\tau$ is $P \in [0,1]$, and the attempted growth event is successful with probability $\newhat{P}\le P$. Again, the difference between $P$ and $\newhat{P}$ is caused by local crowding effects, and since this agent is uncrowded we have $\newhat{P} = P$.  If the attempted growth event is successful, there are two possible outcomes. First, the growth event is a birth event.  In this case a daughter agent is placed at a randomly-chosen vacant site within the set of nearest-neighbour sites with probability $\newhat{P}$. As there are six vacant neighbour sites, the probability of placing a daughter agent at the target site, highlighted in green, is $\newhat{P}/6$. Second, the growth event is a death event, and the agent is removed from the lattice, with probability $\newhat{P}$.  The distinction between the birth and death events is governed by the sign of the \textit{growth crowding function}, $F$, which will be explained later.

To illustrate how crowding effects are incorporated into the movement component of the model, we now consider the schematic in Figure~\ref{fig:2}(c), where the agent at site~$\mathbf{s}$ is surrounded by two agents, highlighted in purple. The probability of attempting to move is $M \in [0,1]$, and the attempted movement event is successful with probability $\newhat{M}=MG(K_{\textbf{s}}^{(\textls{m})})$. Here, $K_{\textbf{s}}^{(\textls{m})}$ is a measure of the local density of site~$\mathbf{s}$, and $G(K_{\textbf{s}}^{(\textls{m})})\in[0,1]$ is the \textit{movement crowding function} that specifies how the local density influences the probability of this agent to undergo a movement event. If this attempt is successful, as there are four vacant neighbour sites, the probability of moving to the target site, highlighted in green, is $\newhat{M}/4$.

Figure~\ref{fig:2}(d) illustrates how crowding effects are incorporated into the growth component of the model, where the agent at site~$\mathbf{s}$ is surrounded by two agents. The probability of attempting to grow is $P \in [0,1]$, and the attempted growth is successful with probability $\newhat{P}=P|F(K_{\textbf{s}}^{(\textls{g})})|$. Here, $K_{\textbf{s}}^{(\textls{g})}$ is again a measure of the local density of site~$\mathbf{s}$ and the function $F(K_{\textbf{s}}^{(\textls{g})})\in[-1,1]$ is called the growth crowding function that specifies how the local density influences the probability of this agent to undergo a growth event. If this attempt is successful, there are two possible outcomes reflected by the sign of $F$. If $F>0$, the growth event is a birth event, and a daughter agent is placed at a randomly-chosen vacant site with probability $\newhat{P}$. As there are four vacant neighbour sites, the probability of placing a daughter agent at the vacant target site, highlighted in green, is $\newhat{P}/4$. Second, if $F<0$, the growth event is a death event, and the agent is removed from the lattice with probability $\newhat{P}$. The special case where $F=0$ leads to neither a birth or death event.

A key feature of our model is in the way that the local density about each site affects movement and growth events through the movement and growth crowding functions. To describe this we take $\mathcal{N}_{r}\{\mathbf{s}\}$ to denote the set of neighbouring sites around site~$\mathbf{s}$, where $r\ge1$ is the integer number of concentric rings of sites surrounding site~$\mathbf{s}$, so that $|\mathcal{N}_{r}|=3r(r+1)$ ~\citep{jin2016stochastic,Fadai2020Proc}. The probability that any potential movement or growth event is successful depends upon the crowdedness of the local region surrounding site~$\mathbf{s}$. We count neighbouring agents in $\mathcal{N}_r$, and consider
\begin{equation}
    \label{template0}
    {K}_{\mathbf{s}}(r)=\dfrac{1}{\lvert{\mathcal{N}_r}\rvert}\sum_{\mathbf{s}'\in \mathcal{N}_r\{\mathbf{s}\}}{C}_{\mathbf{s}'}\in [0,1],
\end{equation}
as a simple measure of the crowdedness of the local region surrounding site~$\mathbf{s}$. While in Figure~\ref{fig:2} we explain the model with $r=1$ and  $|\mathcal{N}_{1}|=6$, it is possible to use different-sized templates, depending on the choice of $r$.  Sometimes it is useful to use different-sized templates for the movement and growth mechanisms.  For example, \citet{simpson2010cell} argues that cell movement can be modelled using a nearest-neighbour random walk with $r=1$, whereas cell proliferation often involves non nearest-neighbour interactions since daughter cells are often deposited several cell diameters away from the location of the mother cell. This argument is supported by experimental images of cell proliferation where careful examination of timelapse movies show that daughter cells are often generated some distance from the mother cell \citep{druckenbrod2005pattern}. To simulate such dynamics, \citet{simpson2010cell} introduce proliferation mechanisms where daughter agents are placed up to four lattice sites away from the mother agent to faithfully capture this biological detail into their model. This would be similar to setting $r=1$ for movement and $r=4$ for growth in our model. It is thus convenient for us to make a notational distinction between the size of the templates for motility and growth. Therefore, we denote the motility template as $K_\mathbf{s}^{(\textls{m})} = K_\mathbf{s}(r')$ and the growth template as $K_\mathbf{s}^{(\textls{g})} = K_\mathbf{s}(r'')$ where $r'\ge 1$ and $r''\ge 1$ are two, potentially different, positive integers.

\begin{landscape}
\begin{figure}
\vspace{-0.6cm}
\centering
\includegraphics[width=1.4\textwidth]{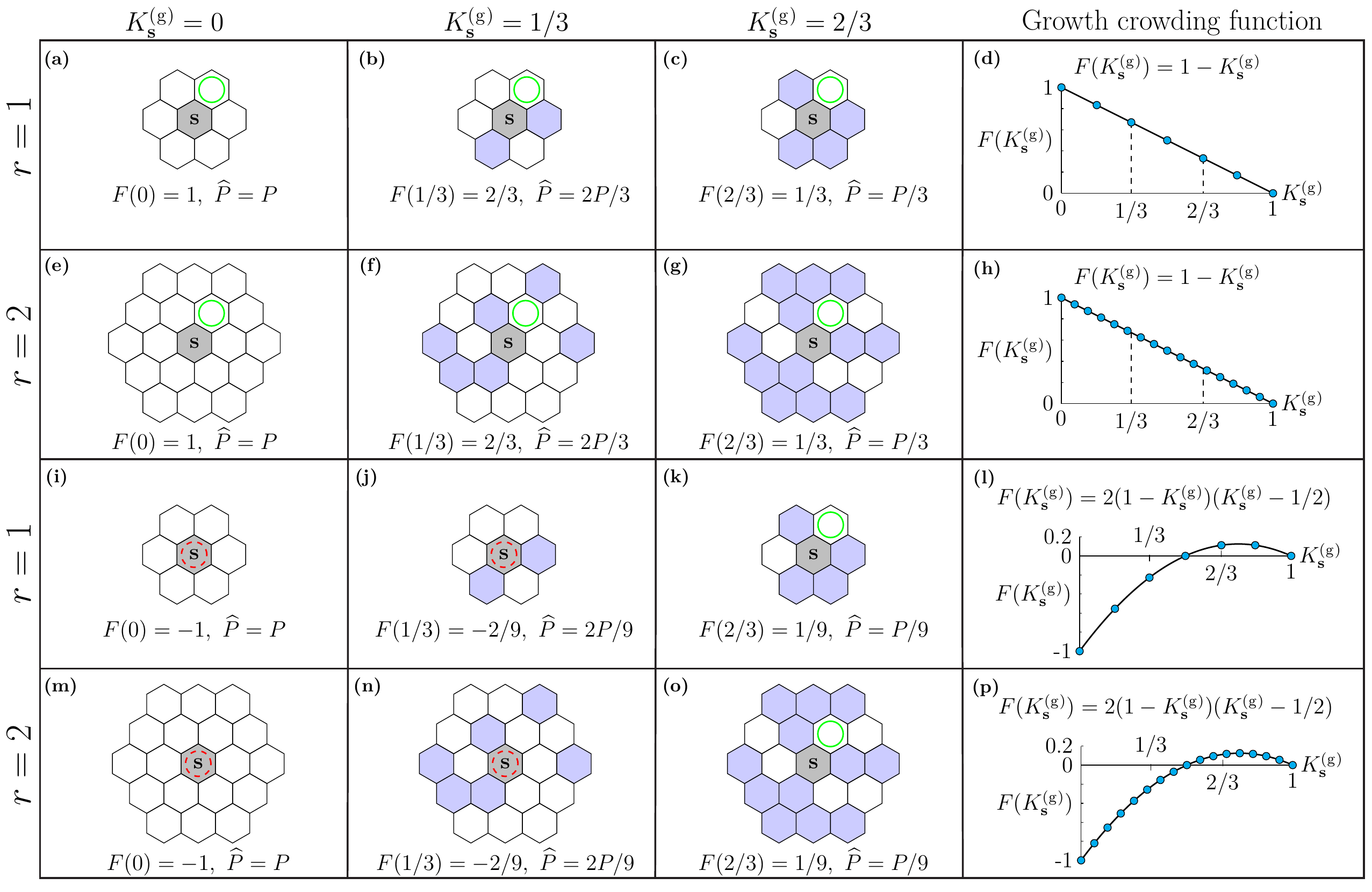}
\singlespace\caption{\textbf{Growth mechanisms with different-sized spatial templates and growth crowding functions}. In each lattice fragment  site~$\mathbf{s}$ is shaded grey,  occupied sites within the template are shaded blue, and vacant sites within the template are unshaded (white). Each subfigure shows a potential outcome for an agent at site~$\mathbf{s}$. The crowdedness of $\mathcal{N}_1$ is shown in (a)--(c) and (i)--(k). The crowdedness of  $\mathcal{N}_2$ is shown in (e)--(g) and (m)--(o). The agent at site~$\mathbf{s}$ can undergo a birth event when $F>0$ as in (a)--(c), (e)--(g), (k) and (o). In contrast the agent at site~$\mathbf{s}$ can undergo a death event when $F<0$ as in (i), (j), (m) and (n). The solid green circles represent the target site for the placement of a daughter agent during a successful proliferation event, and the dashed red circles indicate the location of agents that can undergo a death event.}
\label{fig:3} 
\end{figure}
\end{landscape}

\begin{stretchpars}
We now describe the details of how crowding effects and different-sized spatial templates are incorporated into the growth component of the model with reference to the schematic illustration in Figure~\ref{fig:3}. Note that this figure only indicates the potential growth events without the indication of any movement events. In Figures~\ref{fig:3}(a)--(c),  crowding of the agent at site~$\mathbf{s}$ is measured using a nearest-neighbour template with $r=1$ and the growth crowding function $F(K_\mathbf{s}^{(\textls{g})})=1-K_\mathbf{s}^{(\textls{g})}$, as given in Figure~\ref{fig:3}(d). The probability of undergoing a birth event is $\newhat{P}=P|F(K_{\textbf{s}}^{(\textls{g})})|$. In Figure~\ref{fig:3}(a) where $K_{\textbf{s}}^{(\textls{g})}=0$, we have  $F(0)=1$ and $\newhat{P}=P$. As there are six vacant sites in $\mathcal{N}_1$, the probability of placing a daughter agent at the target site, highlighted in green, is $\newhat{P}/6$. In Figure~\ref{fig:3}(b), where the agent at site~$\mathbf{s}$ is surrounded by two neighbour agents, the probability of undergoing a birth event is $\newhat{P}=2P/3$, since $K_{\textbf{s}}^{(\textls{g})}=1/3$ and $F(1/3)=2/3$. As there are four vacant sites in $\mathcal{N}_1$, the probability of placing a daughter agent at the target site is $\newhat{P}/4$. Similarly, in Figure~\ref{fig:3}(c), we have $\newhat{P}=P/3$ as $K_\mathbf{s}^{(\textls{g})}=2/3$
\end{stretchpars}

\noindent
and $F(2/3)=1/3$. As there are two vacant sites in $\mathcal{N}_1$, the probability of placing a daughter agent at the target site is $\newhat{P}/2$.

In Figures~\ref{fig:3}(e)--(g) we introduce a non nearest-neighbour growth mechanism by measuring the  crowdedness of the agent at site~$\mathbf{s}$ using a larger spatial template with $r=2$. Therefore, if the agent at $\mathbf{s}$ undergoes a successful birth event, the daughter agent is able to be placed at any vacant site within $\mathcal{N}_2$. The probability of undergoing a birth event is $\newhat{P}=P|F(K_\mathbf{s}^{(\textls{g})})|$, where $F(K_\mathbf{s}^{(\textls{g})})=1-K_\mathbf{s}^{(\textls{g})}$. For the agent in Figure~\ref{fig:3}(e) where $K_\mathbf{s}^{(\textls{g})}=0$ and $F(0)=1$, we have $\newhat{P}=P$. In this configuration there are $18$ vacant sites in $\mathcal{N}_2$ and the probability of placing a daughter agent at the target site, highlighted in green, is $\newhat{P}/18$. In Figure~\ref{fig:3}(f), where the agent at site~$\mathbf{s}$ is surrounded by six neighbour agents, the probability of undergoing a birth event is $\newhat{P}=2P/3$, as $K_\mathbf{s}^{(\textls{g})}=1/3$ and
$F(1/3)=2/3$. Since there are $12$ vacant sites in $\mathcal{N}_2$, the probability of placing a daughter agent at the target site is $\newhat{P}/12$. Similarly, in Figure~\ref{fig:3}(g), we have $\newhat{P}=P/3$, as $K_\mathbf{s}^{(\textls{g})}=2/3$ and $F(2/3)=1/3$. The probability of placing a daughter agent at the target site is $\newhat{P}/6$. All results in Figures~\ref{fig:3}(a)--(h) consider the simplest linear crowding function  $F(K_\mathbf{s}^{(\textls{g})})=1-K_\mathbf{s}^{(\textls{g})}$, which means that agents do not die in this case, since $F(K_{\mathbf{s}}^{(\textls{g})})\ge0$.

We now choose a nonlinear growth crowding function $F(K_\mathbf{s}^{(\textls{g})})=2(1-K_\mathbf{s}^{(\textls{g})})(K_\mathbf{s}^{(\textls{g})}-1/2)$ that can take on both positive and negative values, as shown in Figure~\ref{fig:3}(l). In this case we make a distinction between a birth event when $F(K_\mathbf{s}^{(\textls{g})}) > 0$, a death event when $F(K_\mathbf{s}^{(\textls{g})})<0$, and no event when $F(K_\mathbf{s}^{(\textls{g})})=0$. We first consider a nearest-neighbour template with $r=1$  in Figures~\ref{fig:3}(i)--(k). In Figure~\ref{fig:3}(i), the agent at site~$\mathbf{s}$ dies with probability $\newhat{P}=P|F(K_\mathbf{s}^{(\textls{g})})|$. Here, $K_\mathbf{s}^{(\textls{g})}=0$ and $F(0)=-1$, thus $\newhat{P}=P$. In Figure~\ref{fig:3}(j) the agent at site~$\mathbf{s}$ dies with probability $\newhat{P}=2P/9$ as $K_\mathbf{s}^{(\textls{g})}=1/3$ and $F(1/3)=-2/9$. In Figure~\ref{fig:3}(k) the agent at site~$\mathbf{s}$ undergoes a birth event with probability $\newhat{P}=P/9$ as $K_\mathbf{s}^{(\textls{g})}=2/3$ and $F(2/3)=1/9$. As there are two vacant sites in $\mathcal{N}_1$, the probability of placing a daughter agent at the target site is $\newhat{P}/2$. 

Finally, we consider a larger template with $\mathcal{N}_2$ in Figures~\ref{fig:3}(m)--(o). In Figure~\ref{fig:3}(m), the agent at site~$\mathbf{s}$ dies with probability $\newhat{P}=P|F(K_\mathbf{s}^{(\textls{g})})|$, where $F(K_\mathbf{s}^{(\textls{g})})=2(1-K_\mathbf{s}^{(\textls{g})})(K_\mathbf{s}^{(\textls{g})}-1/2)$. Here $K_\mathbf{s}^{(\textls{g})}=0$ and $F(0)=-1$, thus $\newhat{P}=P$. In Figure~\ref{fig:3}(n), the agent at site~$\mathbf{s}$ dies with probability $\newhat{P}=2P/9$ as $K_\mathbf{s}^{(\textls{g})}=1/3$ and $F(1/3)=-2/9$. In Figure~\ref{fig:3}(o) the agent at site~$\mathbf{s}$ undergoes a birth event with probability $\newhat{P}=P/9$ as $K_\mathbf{s}^{(\textls{g})}=2/3$ and $F(2/3)=1/9$. As there are six vacant sites in $\mathcal{N}_2$, the probability of placing a daughter agent at the target site is $\newhat{P}/6$.

The movement crowding function, $G(K_\mathbf{s}^{(\textls{m})})$, is incorporated into the model in a similar way as the growth crowding function except that it is always non-negative, $G(K_\mathbf{s}^{(\textls{m})})\in[0,1]$. In this section we have sought to describe the discrete mechanism as clearly as possible with the use of Figure~\ref{fig:2} and Figure~\ref{fig:3}.  For the remainder of this work we focus on results where we set $r=1$ for movement and $r=4$ for growth. Other choices of $r$ can be implemented using the software available on    \href{https://github.com/oneflyli/Yifei2020Dimensionality}{GitHub}.

\section{Continuum limit}
\label{sec:3}
In this section we derive the mean-field continuum limit of the discrete model. The \textit{averaged} occupancy of site~$\mathbf{s}$, constructed from  $V$ identically-prepared realisations of the discrete model, can be written as
\begin{equation}
    \label{averageoccupancy}
    \newbar{C}_{\mathbf{s}}=\frac{1}{V}\sum_{v=1}^{V}C_\mathbf{s}^{(v)}(t),
\end{equation}
where ${C}_{\mathbf{s}}^{(v)}(t) \in \{0,1\}$ is the binary occupancy of site~$\mathbf{s}$ at time $t$ in the $v$th identically-prepared realisation of the discrete model.  We note that $\newbar{C}_{\mathbf{s}}\in[0,1]$, and is a function of time, $t$, but we suppress this dependence for notational convenience. Similarly, the \textit{averaged} occupancy of $\mathcal{N}_r\{\mathbf{s}\}$, again constructed from $V$ identically-prepared realisations, is given by
\begin{equation}
    \label{template0'}
    \newbar{K}_{\mathbf{s}}(r)=\dfrac{1}{\lvert{\mathcal{N}_r}\rvert}\sum_{\mathbf{s}'\in \mathcal{N}_r\{\mathbf{s}\}}\newbar{C}_{\mathbf{s}'}.
\end{equation}
As we use a nearest-neighbour template, $r=1$, for movement, and a larger template, $r=4$, for growth, we denote the averaged occupancy of sites for potential movement events as $\newbar{K}_\mathbf{s}^{(\textls{m})}$, and the averaged occupancy of sites for potential growth events as $\newbar{K}_\mathbf{s}^{(\textls{g})}$. 

To arrive at an approximate continuum limit description, we start by writing down an expression for the expected change in occupancy of site~$\mathbf{s}$ during the time interval from $t$ to $t+\tau$, 
\begin{equation}
\label{delta1}
     \begin{aligned}
     \delta(\newbar{C}_{\mathbf{s}})=&\overbrace{\frac{M}{\lvert\mathcal{N}_{1}\rvert}(1-\newbar{C}_{\mathbf{s}})\sum_{\mathbf{s}'\in \mathcal{N}_{1}\{\mathbf{s}\}}\newbar{C}_{\mathbf{s}'}\frac{G(\newbar{K}_{\mathbf{s}'}^{(\textls{m})})}{1-\newbar{K}_{\mathbf{s}'}^{(\textls{m})}}}^\text{\normalsize movement events into $\mathbf{s}$}\quad -\quad \overbrace{\vphantom{\frac{G(\newbar{K}_{\mathbf{s}}^{(\textls{m})})}{1-\newbar{K}_{\mathbf{s}}^{(\textls{m})}}}M\newbar{C}_\mathbf{s}G(\newbar{K}_{\mathbf{s}}^{(\textls{m})})}^\text{\normalsize movement events out of $\mathbf{s}$}\\
          &+\underbrace{\frac{P}{\lvert\mathcal{N}_{4}\rvert}(1-\newbar{C}_{\mathbf{s}})\sum_{\mathbf{s}'\in \mathcal{N}_{4}\{\mathbf{s}\}}\mathbbm{H}(F(\newbar{K}_{\mathbf{s}'}^{(\textls{g})}))\newbar{C}_{\mathbf{s}'}\frac{F(\newbar{K}_{\mathbf{s}'}^{(\textls{g})})}{1-\newbar{K}_{\mathbf{s}'}^{(\textls{g})}}}_\text{\normalsize birth events: place new agents onto $\mathbf{s}$} 
          \\
          &-\underbrace{ (1-\mathbbm{H}(F(\newbar{K}_{\mathbf{s}}^{(\textls{g})})){P\newbar{C}_\mathbf{s}}F(\newbar{K}_{\mathbf{s}}^{(\textls{g})})}_\text{\normalsize death events: remove agent from $\mathbf{s}$},
     \end{aligned}
\end{equation}
where $\mathbbm{H}$ is the Heaviside step function. Each term in Equation~\eqref{delta1} has a relatively simple physical interpretation. The first term on the right hand side of Equation~\eqref{delta1} represents the change in occupancy of site~$\mathbf{s}$ owing to the expected movement of agents in $\mathcal{N}_{1}\{\mathbf{s}\}$ into site~$\mathbf{s}$. The factor $1/(1-\newbar{K}_\mathbf{s}^{(\textls{m})})$ accounts for the choice of the target site in $\mathcal{N}_1$ being randomly selected from the available vacant sites. The second term on the right hand side of Equation~\eqref{delta1} represents the change in occupancy of site~$\mathbf{s}$ owing to the expected movement of agents out of site~$\mathbf{s}$. The third term on the right hand side of Equation~\eqref{delta1} represents the change in occupancy owing to the expected birth events of agents in $\mathcal{N}_{4}\{\mathbf{s}\}$ that would place daughter agents onto site~$\mathbf{s}$, where $F(\newbar{K}_{\mathbf{s}}^{(\textls{g})})>0$. Again, the factor $1/(1-\newbar{K}_\mathbf{s}^{(\textls{g})})$  accounts for the choice of the target site in $\mathcal{N}_4$ being randomly selected from the available vacant sites. The last term on the right hand side of Equation~\eqref{delta1} represents the expected change in occupancy owing to agent death at site~$\mathbf{s}$, when  $F(\newbar{K}_{\mathbf{s}}^{(\textls{g})})<0$. Note that this approximate conservation statement \eqref{delta1} makes use of the mean-field assumption, whereby the occupancy status of  lattice sites are taken to be independent~\citep{Baker2010}.

To derive the continuum limit we replace $\newbar{C}_\mathbf{s}$ with a continuous function, $C(x,y,t)$, and expand each term in Equation~\eqref{delta1} in a Taylor series
about site~$\mathbf{s}$, and truncate terms of $\mathcal{O}(\Delta^3)$. Subsequently, we divide both sides of the resulting expression by  $\tau$ and evaluate the resulting expressions in the limit $\Delta \to0$ and $\tau\to0$ jointly, with the ratio of $\Delta^2/\tau$ held constant \citep{hughes1995random}. This leads to the following nonlinear RDE,
\begin{equation}
    \label{PDE_general_2D_general}
    \frac{\partial C(x,y,t)}{\partial t}=D_0\nabla\cdot\left(D(C)\nabla C(x,y,t)\right)+\lambda C(x,y,t)F(C),
\end{equation}
where
\begin{equation}
\label{diffusionterm}
    D(C)=C\frac{\textrm{d}G(C)}{\textrm{d}C}+\frac{1+C}{1-C}G(C),
\end{equation}
and 
\begin{equation}
\label{limitparameter1}
    D_0=\frac{M}{4}\lim_{\Delta, \tau\to 0}
    \frac{\Delta^2}{\tau},\quad \lambda=\lim_{\tau\to0}\frac{P}{\tau}.
\end{equation}
Here, $D_0$ is the free-agent diffusivity, $D(C)$ is a nonlinear diffusivity function that relates to the movement crowding function $G(C)$, and $\lambda$ is the rate coefficient associated with the source term that is related to the  growth crowding function $F(C)$.  To obtain a well-defined continuum limit we require that $P = \mathcal{O}(\tau)$ \citep{simpson2010cell}.  The algebraic details required to arrive at the continuum limit are outlined in the Supplementary Material.

For all simulations in this work we use $\Delta=\tau=1$, giving $D_0=M/4$ and $\lambda=P$. This is equivalent to working in a non-dimensional framework \citep{simpson2010cell}.  If the model is to be applied to a particular dimensional problem, then $\Delta$ and $\tau$ can be re-scaled using appropriate length and time scales. In this non-dimensional framework with $\tau=1$, we satisfy the requirement that $P =\mathcal{O}(\tau)$ by ensuring $P/M \ll 1$. The main focus of this work is on the role of the growth mechanism, and the question of whether the population survives or goes extinct. We therefore set 
$
    G(C)=1-C
$
leading to $D(C)=1$. This means that the nonlinear diffusion term in Equation~\eqref{PDE_general_2D_general} simplifies to a linear diffusion term, giving 
\begin{equation}
    \label{PDE_general_2D}
    \frac{\partial C(x,y,t)}{\partial t}=D_0\nabla^2C(x,y,t)+\lambda C(x,y,t)F(C).
\end{equation}
We note that Equation~\eqref{PDE_general_2D} has been studied extensively in applications involving the spatial spread of invasive species, such as the works of \citet{Fisher1937,Skellam1951,fife1979long,lewis1993allee} and \citet{Hastings2005}. Some previous models consider a logistic-type source term  \citep{Fisher1937}, while others consider Allee-type bistable source term \citep{Sewalt2016}. Under these conditions many results have been established.  For example, if we consider Equation \eqref{PDE_general_2D} on a one-dimensional infinite domain, it is well known that this model supports travelling wave solutions for both logistic \citep{Fisher1937} and bistable \citep{fife1979long} source terms.  In this work, however, we take a different perspective by studying Equation~\eqref{PDE_general_2D} on a finite domain and so the question of analysing travelling wave solutions is not our focus. Moreover, although \citet{lewis1993allee} give a critical radius of a radially symmetric distribution so that the initial distribution converges to an expanding wave in an infinite domain, their analysis is valid under the assumption that the time scale of 
growth is much faster than the time scale of migration, which corresponds to $P/M \gg 1$ in our framework.  Our discrete model does not have any such restriction and can be implemented for any $M \in [0,1]$ and any $P \in [0,1]$.  In contrast, our continuum model requires $P/M \ll 1$ to correspond to the discrete model, and we will explore the consequences of these differences in our results.

\begin{figure}[]
\centering
\includegraphics[]{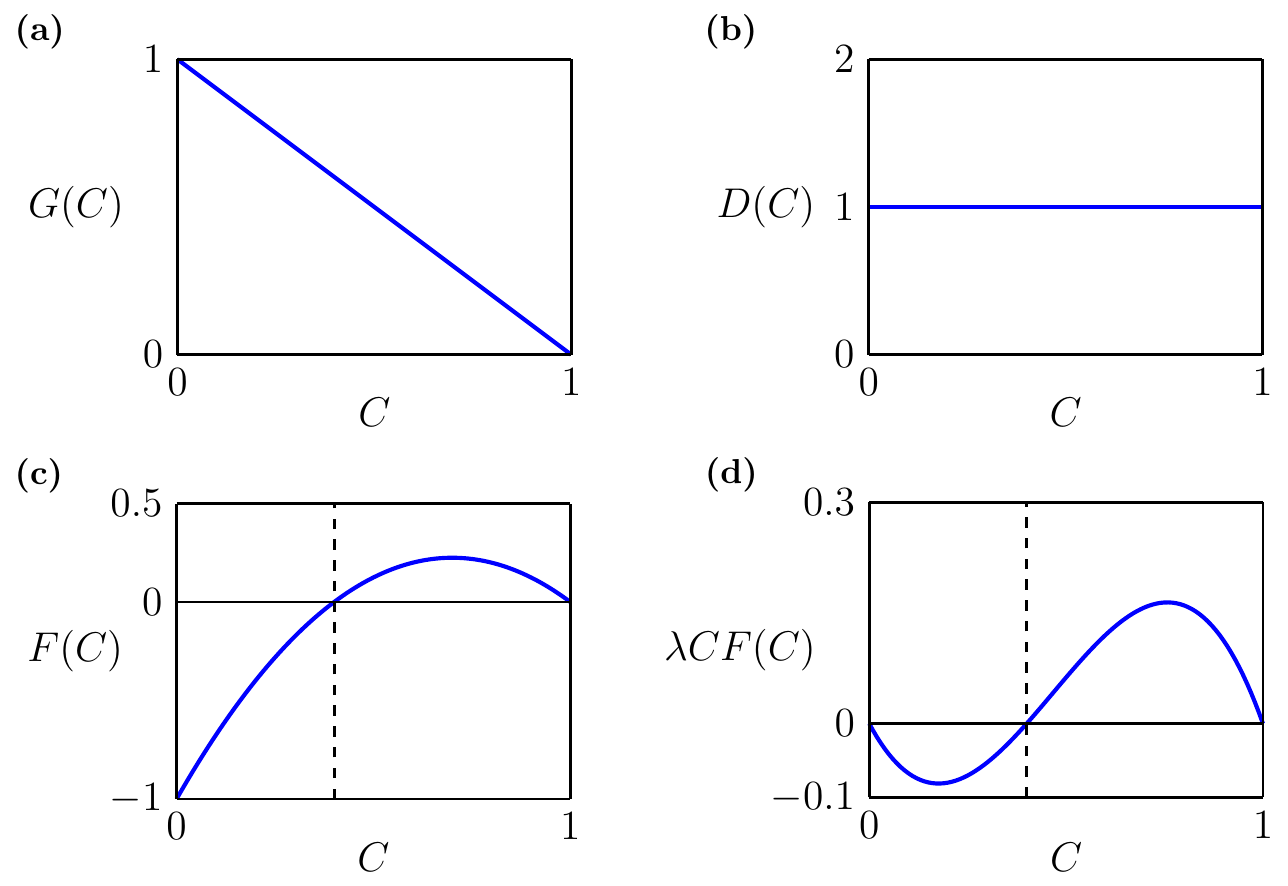}
\singlespace\caption{\textbf{Specific crowding functions used in this work.} (a)--(b) Setting $G(C)=1-C$ for the movement crowding function leads to linear diffusion,  $D(C)=1$. (c)--(d) Setting $F(C) = 5(1-C)(C-2/5)/2$ for the growth crowding function with $\lambda=P=1$ leads to $\lambda CF(C) = 5C(1-C)(C-2/5)/2$. The dashed lines in (c)--(d) relate to the Allee threshold, $A=2/5$.}
\label{fig:4} 
\end{figure}

In the rest of this work we choose
\begin{equation}
    \label{reactionterm_1}
    F(C)=a(1-C)(C-A),\quad \text{with}\quad  a=\dfrac{5}{2},\quad A=\dfrac{2}{5},
\end{equation}
since this leads to the canonical cubic source term $\lambda CF(C)$ associated with Allee kinetics. In particular, we set $A=2/5$ so that this choice of $F(C)$ can be used to represent birth events where $C>2/5$ and death events where $C<2/5$, see Figure~\ref{fig:4}(d). We further set $a=5/2$ leading to $F(0)=-1$, so that attempted death events for an isolated agent, where $C=0$, are always successful. 

In summary, our discrete model requires the specification of two crowding functions: $G(C)$ and $F(C)$. These crowding functions are related to macroscopic quantities in the associated RDE model.  In particular, $G(C)$ is related to a nonlinear diffusivity function, $D(C)$, and $F(C)$ is related to a nonlinear source term $\lambda CF(C)$.  Figure~\ref{fig:4} shows the relationship between these functions for our choice of $G(C)$ and $F(C)$.

\section{Initial distributions and simulation data}
\label{sec:4}
In this section we consider the three initial distributions shown in Figure~\ref{fig:1} with $L=100$, and we introduce the corresponding continuous descriptions. In general, each of the initial distribution shown in Figure~\ref{fig:1} can be written as 
\begin{equation}
    \label{initialdistribution}
    C(x,y,0)=\left\{\begin{aligned}
    &B,\quad (x,y)\in\mathcal{H},\\
    &0,\quad \text{elsewhere},
    \end{aligned}\right.
\end{equation}
where $\mathcal{H}$ is the region in which individuals are distributed at density $B\in(0,1]$. For the discrete model, we randomly distribute a fixed number of agents on $\mathcal{H}$ so that the averaged density across $\mathcal{H}$ is $B$. For example, all agents in the discrete model are closely packed together if $B=1$. In contrast, for the continuum model, the density is $B$ at each position in $\mathcal{H}$.

For the three initial distributions in Figure~\ref{fig:1} we will report data from the stochastic model in the following way. We denote the averaged occupancy of site~$\mathbf{s}$ in $V$ identically-prepared simulations as
\begin{equation}
\label{approximate_cxyt}
    \left<C(x,y,t)\right>=\frac{1}{V}\sum_{v=1}^{V}C^{(v)}(i,j,n),
\end{equation}
where we note that the average denoted by the angular parenthesis is taken in the same way as the average in Equation~\eqref{averageoccupancy}.  Here, site $\mathbf{s}$, indexed by $i$ and $j$, are related to position, $(x,y)$ via Equation~\eqref{xyijrelation}. The averaged occupancy $\left<C(x,y,t)\right>$ is a measure of the local density at location $(x,y)$, and time $t=n\tau$ after the $n$th time step in the stochastic discrete model. Although \eqref{approximate_cxyt} describes the averaged occupancy of any distribution, there are more concise forms for the vertical strip distributions in Figure~\ref{fig:1}(b). As the initial occupancy is independent of the vertical position, we denote the averaged occupancy of any site as
\begin{equation}
\label{approximate_cxt}
    \langle C(x,t) \rangle=\frac{1}{VJ}\sum_{v=1}^{V}\sum_{j=1}^{J}C^{(v)}(i,j,n),
\end{equation}
which is a measure of the density at location $x$ and at time $t=n\tau$. Note that, as we will show through simulation, the density of agents remains independent of the vertical position for all $t > 0$ because we use periodic boundary conditions. Similarly, for simulations relating to the well-mixed initial distribution as in Figure~\ref{fig:1}(a), where the initial density is independent of position, we denote the averaged occupancy of any site as
\begin{equation}
    \label{approximate_ct}
    \langle C(t) \rangle=\frac{1}{VIJ}\sum_{v=1}^{V}\sum_{j=1}^{J}\sum_{i=1}^{I}C^{(v)}(i,j,n),
\end{equation}
which is a measure of the total population density at time $t=n\tau$.  As we will show through simulation, in this case the density of agents remains independent of position for all $t > 0$. The total population density $\langle C(t) \rangle$ is also useful to describe simulations starting from the square and vertical strip initial distributions. In summary, data from the discrete models can be summarised by calculating $\langle C(x,y,t)\rangle$, $\langle C(x,t)\rangle$, and $\langle C(t)\rangle$.

\begin{figure}[]
\centering
\includegraphics[width=\textwidth]{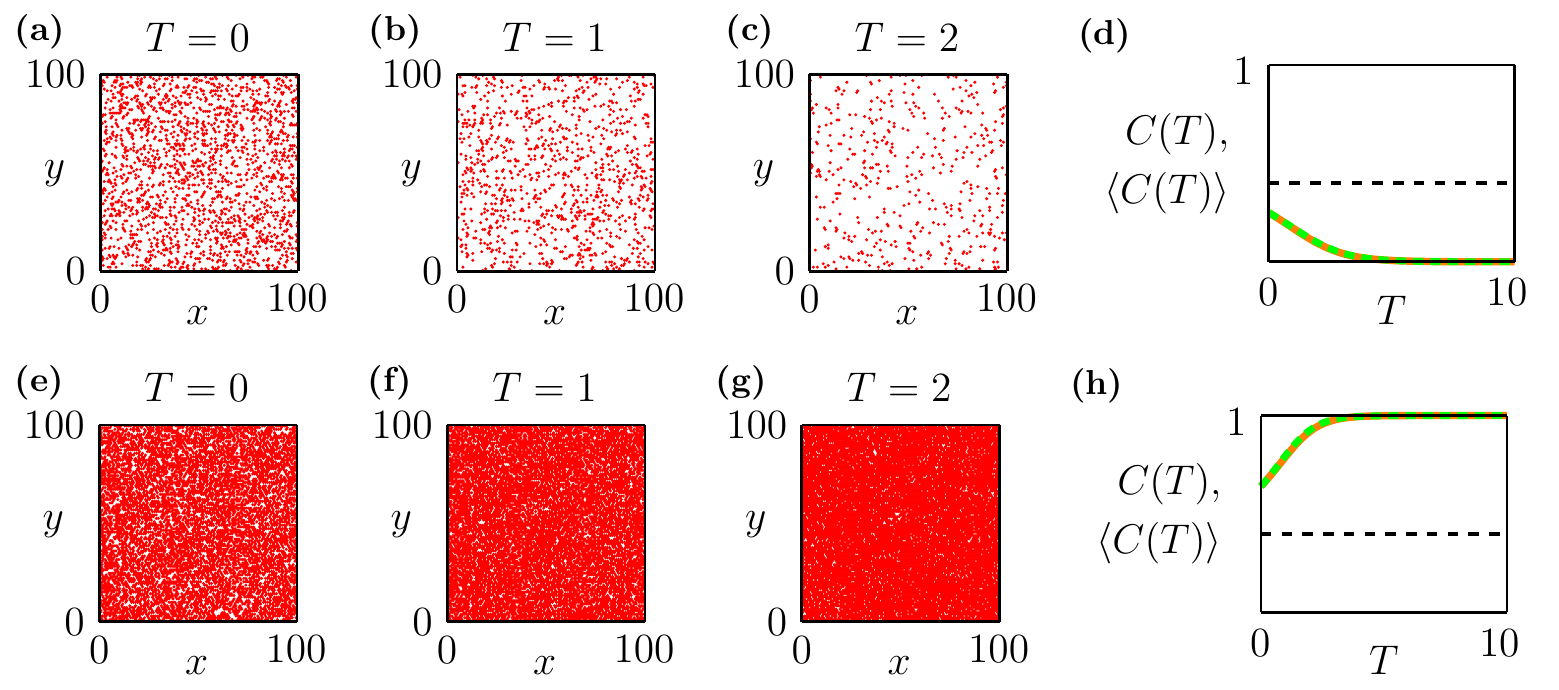}
\singlespace\caption{\textbf{Comparison of data from the discrete model with the solution of the continuum model for the well-mixed initial distribution.} (a)--(c) Snapshots of discrete simulations at time $T=\lambda t=0,1,2$. At $T=0$ a fixed number of agents are randomly distributed on the lattice so that $\langle C(0)\rangle=0.25$. (d) $\langle C(T)\rangle$ (solid orange) and $C(T)$ (dashed green). (e)--(g) Snapshots of discrete simulations at time $T=0,1,2$ with $\langle C(0)\rangle=0.64$. (h) $\langle C(T)\rangle$ (solid orange) and $C(T)$ (dashed green). The dashed black horizontal lines in (d) and (h) are the Allee threshold, $A=0.4$.}
\label{fig:5} 
\end{figure}

For the well-mixed initial distribution, as shown in Figure~\ref{fig:1}(a), Equation~\eqref{PDE_general_2D} simplifies to
\begin{equation}
    \label{PDE_general_0D}
    \frac{\textrm{d} C(t)}{\textrm{d} t}=\lambda C(t) F(C), 
\end{equation}
where $C(t)$ represents the total density of the population \citep{simpson2010cell}. This separable ODE can be solved to give an implicit solution for our choice of $F(C)$. Results in Figure~\ref{fig:5} compare the discrete and continuum solutions for the well-mixed initial distribution. In Figure~\ref{fig:5}(a), a fixed number of agents are randomly distributed in the entire domain at $T=\lambda t=0$, leading to $\langle C(0)\rangle=0.25$. Figures~\ref{fig:5}(b)--(c) show discrete snapshots as the population evolves with $M=1$ and $P=1/1000$, leading to $\lambda=1/1000$ in \eqref{PDE_general_0D}. We superimpose the solution of Equation~\eqref{PDE_general_0D} with averaged data from the discrete model in Figure~\ref{fig:5}(d). The continuum model gives a good approximation to the averaged discrete data, and in this case we see that the population becomes extinct. Note that we generate $V=40$ identically-prepared realisations to obtain $\langle C(T)\rangle$ in Figure~\ref{fig:5}(d) and (h). The estimated robustness of the averaged data is presented in the Supplementary Material.

We now consider the exact same discrete mechanism with a larger initial number of agents giving $\langle C(0)\rangle=0.64$ in Figure~\ref{fig:5}(e). Figures~\ref{fig:5}(f)--(g) again show discrete snapshots as the population evolves, and we observe that $C(T)$ approximates $\langle C(T)\rangle$ well in Figure~\ref{fig:5}(h). In this case the population survives and grows to reach the maximum density.

For the vertical strip initial distribution, as shown in Figure~\ref{fig:1}(b), Equation~\eqref{PDE_general_2D} simplifies to
\begin{equation}
    \label{PDE_general_1D}
    \frac{\partial C(x,t)}{\partial t}=D_0\frac{\partial^2 C(x,t)}{\partial x^2}+\lambda C(x,t) F(C),
\end{equation}
where $C(x,t)$ represents the column-averaged density of agents \citep{simpson2010cell}. An extensive discussion and exploration of the implications of simplifying the two-dimensional nonlinear RDE into this simpler one-dimensional RDE is given in \citet{Simpson2009depth}.  Given a numerical solution of Equation~\eqref{PDE_general_1D}, as outlined in the Supplementary Material, we compute
\begin{equation}
    \label{totaldensity_1D}
    \mathcal{C}(t)=\frac{1}{L}\int_0^{L}C(x,t)\ \textrm{d}x,
\end{equation}
which is the total density of the population in the whole domain, and corresponds to $\left<C(t)\right>$ in the discrete model.

\begin{figure}
\centering
\includegraphics[width=0.9\textwidth]{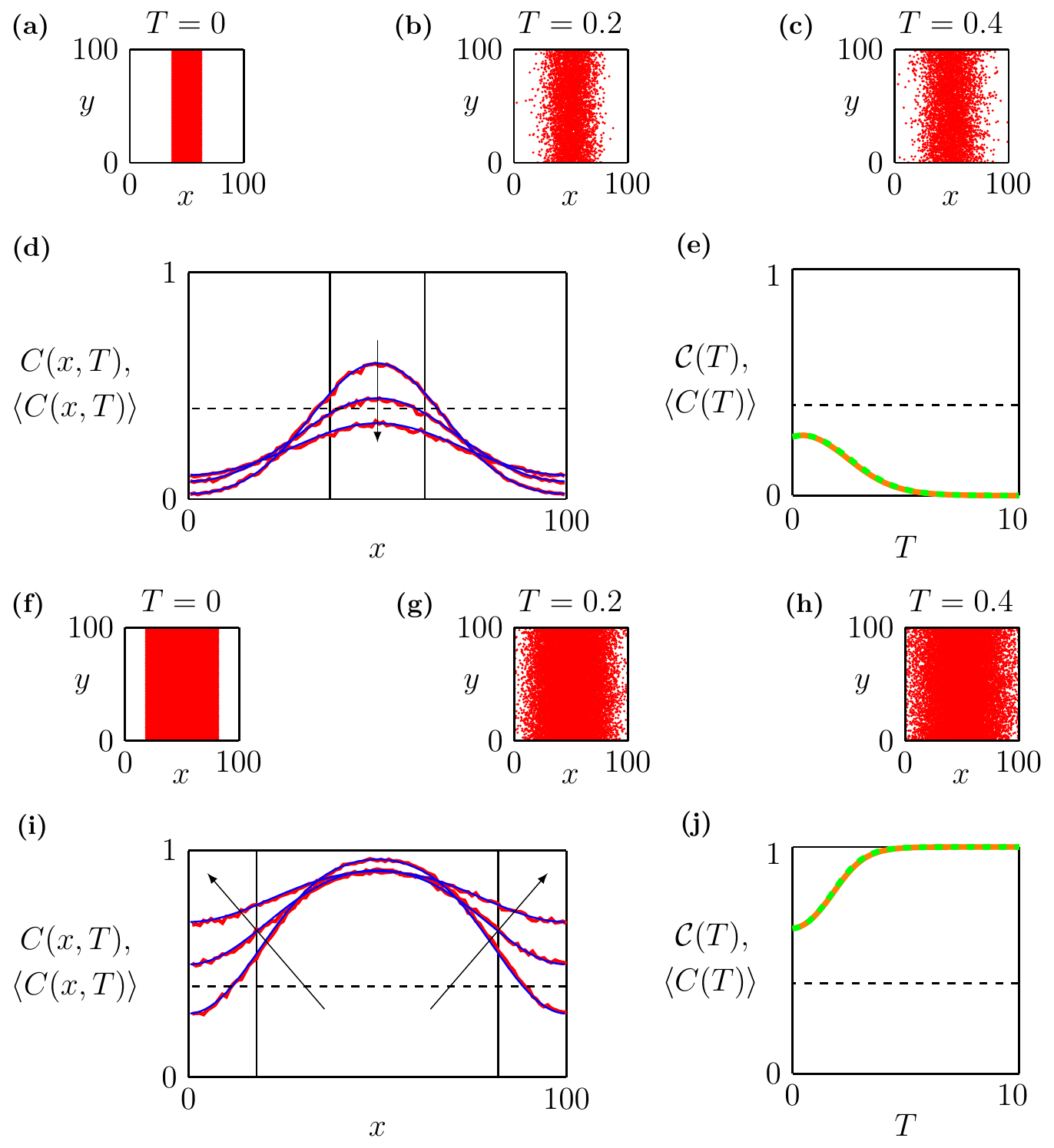}
\singlespace\caption{\textbf{Comparison of data from the discrete model with the solution of the continuum model for the vertical strip initial distribution.} (a) Agents are initially placed within a vertical strip where $x\in[37.5,62.5]$, with $B=1$. (b)--(c) Snapshots from the discrete model at $T=0.1$ and $T=0.2$, respectively. (d) $\langle C(x,T) \rangle$ (red) and $C(x,T)$ (blue) at time $T=0.6,1,2,1.8$. (e) $\langle C(t)\rangle$ (solid orange) and $\mathcal{C}(t)$ (dashed green). (f) Agents are initially placed within a vertical strip where $x\in[18,82]$, with $B=1$. (g)--(h) Snapshots from the discrete model at $T=0.1$ and $T=0.2$, respectively. (i) $\langle C(x,T) \rangle$ (red) and $C(x,T)$ (blue) at time $T=0.6,1.2,1.8$. (j) $\langle C(t)\rangle$ (solid orange) and $\mathcal{C}(t)$ (dashed green). The dashed black horizontal lines in (d), (e), (i) and (j) indicate the Allee threshold, $A=0.4$.  Arrows in (d) and (i) show the direction of increasing time. Note that we generate 40 identically-prepared realisations to obtain $\langle C(x,T)\rangle$ in (d) and (i), and $\langle C(T)\rangle$ in (e) and (j).}
\label{fig:6} 
\end{figure}

Results in Figure~\ref{fig:6} give a comparison between the discrete and continuum solutions for the vertical strip initial distribution. Simulations are performed with $M=1$ and $P=1/1000$, leading to $D_0=1/4$ and $\lambda=1/1000$. The initial distribution in Figure~\ref{fig:6}(a) shows that the central strip of width $25$ is occupied with density $B=1$. Figures~\ref{fig:6}(b)--(c) show snapshots from the discrete model as the population spreads into the domain. Figure~\ref{fig:6}(d) compares the numerical solution of 
Equation~\eqref{PDE_general_1D}, $C(x,T)$, with averaged data from the discrete model, $\langle C(x,T)\rangle$. The evolution of the total population density in the discrete model, $\langle C(T) \rangle$, and in the continuum model, $\mathcal{C}(T)$, is compared in Figure~\ref{fig:6}(e). In all cases the continuum model accurately captures the averaged data from the discrete model, and in this case the population eventually becomes extinct. This is an interesting result given that the initial density in the central strip is greater than the Allee threshold, yet the total population eventually becomes extinct as the migration of individuals reduces the population density locally to below the Allee threshold.

We then consider a second set of discrete-continuum comparisons for precisely the same mechanisms except that the spatial arrangement of the vertical strip initial distribution, shown in Figure~\ref{fig:6}(f), is wider and occupies the central vertical strip of width $64$ with density $B=1$. Figures~\ref{fig:6}(g)--(h) show discrete snapshots as the population spreads. The comparisons between $C(x,T)$ and $\langle C(x,T)\rangle$ in Figure~\ref{fig:6}(i), and between $\mathcal{C}(T)$ and $\langle C(T)\rangle$ in Figure~\ref{fig:6}(j) are excellent. In this case we see that the population eventually grows to reach the maximum density. For the vertical strip initial distribution the same discrete mechanism again leads to different long-term outcomes in Figures~\ref{fig:6}(a)--(e) and Figures~\ref{fig:6}(f)--(j), where the population eventually becomes extinct in the former case, while surviving in the latter case.  The only difference is in the width of the initial population.

For the square initial distribution, to compare averaged data from the discrete model with the solution of the continuum model we solve Equation~\eqref{PDE_general_2D} numerically to give  $C(x,y,t)$. Full details of the numerical methods are presented in the Supplementary Material. Using the numerical solution for $C(x,y,t)$ we calculate
\begin{equation}
    \label{totaldensity_2D}
    \mathcal{C}(t)=\frac{1}{L^2}\int_0^{L}\int_0^{L}C(x,y,t)\ \textrm{d}x\ \textrm{d}y,
\end{equation}
which, again, is the total density of the population in the whole domain.

\begin{figure}[]
\centering
\includegraphics[width=0.95\textwidth]{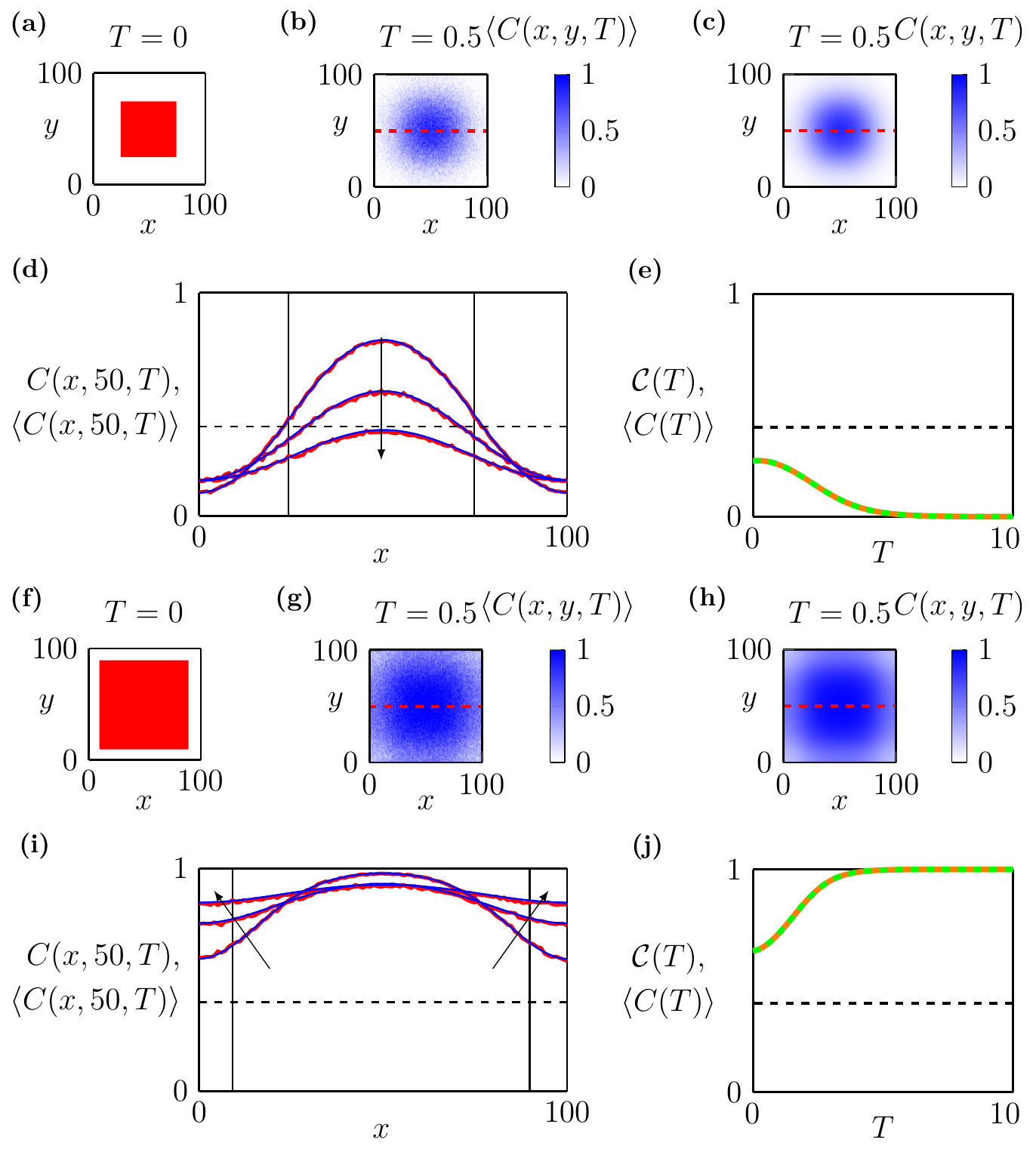}
\singlespace\caption{\textbf{Comparison of data from the discrete model with the solution of the continuum model for the square initial distribution.} (a) Agents are initially located in a square region of size $40\times40$ with $B=1$. (b) $\langle C(x,y,T)\rangle$ at $T=\lambda t=0.5$. (c) $C(x,y,T)$ at $T=\lambda t = 0.5$. (d) $\langle C(x,50,T) \rangle$ (red) and $C(x,50,T)$ (blue) at $T=0.6, 1.2, 1.8$. (e) $\langle C(T) \rangle$ (solid grey) and $\mathcal{C}(T)$ (dashed green). (f) Agents are initially located at a square region of size $80\times80$ with $B=1$. (g) $\langle C(x,y,T)\rangle$ at $T=\lambda t = 0.5$. (h) $C(x,y,T)$ at time $T=\lambda t = 0.5$. (i) $\langle C(x,50,T) \rangle$ (red) and $C(x,50,T)$ (blue) at $T=0.6, 1.2, 1.8$. (j) $\langle C(T) \rangle$ (solid grey) and $\mathcal{C}(T)$ (dashed green). The dashed red lines in (b), (e), (g), (h) indicate the line $y=50$, where we obtain the density along the horizontal direction. The dashed black horizontal lines in (d), (e), (i) and (j) indicate the Allee threshold, $A=0.4$. Arrows in (d) and (i) show the direction of increasing time. Note that we generate 40 identically-prepared realisations to obtain $\langle C(x,y,T)\rangle$ in (b) and (g), and $\langle C(T)\rangle$ in (e) and (j). While we use 4000 identically-prepared realisations to obtain $\langle C(x,50,T)\rangle$ in (d) and (i).}
\label{fig:7} 
\end{figure}

In Figure~\ref{fig:7}, we compare data from the discrete model with numerical solutions of the continuum model for the square initial distribution. Again, simulations are performed with $M=1$ and $P=1/1000$, leading to $D_0=1/4$ and $\lambda=1/1000$. The initial distribution in Figure~\ref{fig:7}(a) shows a square region of size $50\times50$ that is occupied with density $B=1$. Figure~\ref{fig:7}(b) shows a snapshot from the discrete model at $T = \lambda t = 0.5$ where we see the agents spreading into the domain. The numerical solution of Equation~\eqref{PDE_general_2D} in Figure~\ref{fig:7}(c) shows the solution of the continuum model at $T=0.5$. The visual comparison between the spatial arrangement of agents in the discrete model and the density of the profiles in Figure~\ref{fig:7}(b) and Figure~\ref{fig:7}(c) matches well. To make a more quantitative comparison we examine the density along the horizontal dashed lines shown in Figures~\ref{fig:7}(b)--(c) at $y=50$.  Figure~\ref{fig:7}(d) compares the evolution of $C(x,50,T)$ and $\langle C(x,50,T)\rangle$, and we see that the match between the solution of the continuum model and appropriately averaged data from the discrete model is excellent. Finally, in Figure~\ref{fig:7}(e) we compare the averaged total occupancy from the discrete model, $\langle C(T) \rangle$, with  $\mathcal{C}(T)$ from the solution of the continuum model. Again, we see that the discrete-continuum comparison is excellent, and that the continuum model predicts the eventual extinction of this population. Similar to the outcomes from the one-dimensional initial distributions, although the initial density of agents in the central of the domain exceeds the Allee threshold, the migration of individuals reduces the density locally to below the Allee threshold, resulting in extinction.

We now consider a second set of discrete-continuum comparisons for precisely the same mechanisms except that the spatial arrangement of the square initial distribution, shown in Figure~\ref{fig:7}(f), is larger and occupies the central $80 \times 80$ region of the domain. As before, the match between the discrete averaged data and numerical solutions of Equation~\eqref{PDE_general_2D} is excellent in Figures~\ref{fig:7}(g)--(j). In this case the population eventually grows to reach the maximum density.

\begin{figure}[]
\centering
\includegraphics[width=0.85\textwidth]{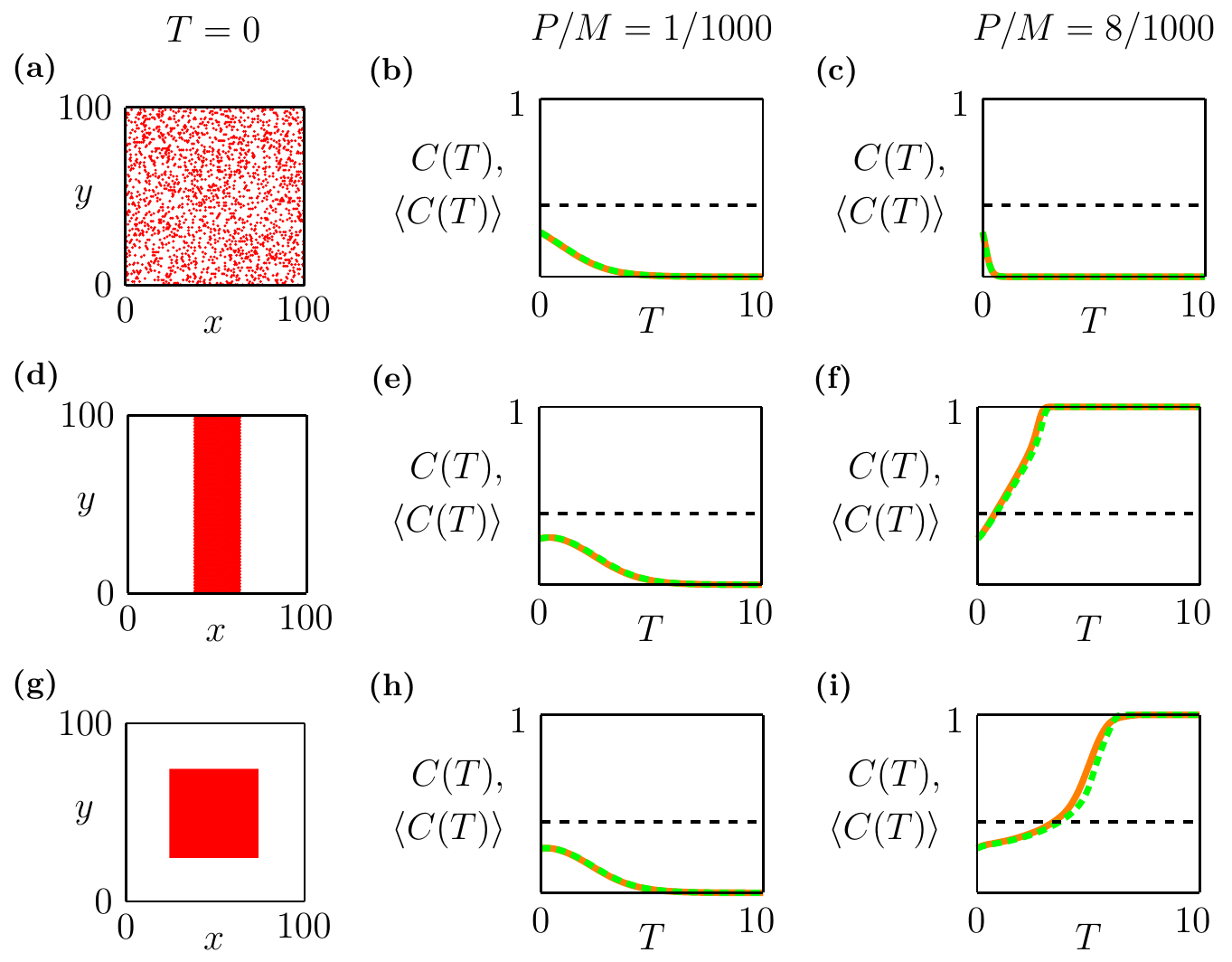}
\singlespace\caption{\textbf{The ratio $P/M$ and the shape of the initial spatial distribution influence the fate of populations.} (a) Well-mixed initial distribution with $\mathcal{C}(0)=0.25$. (b) $\langle C(t)\rangle$ (solid orange) and $\mathcal{C}(t)$ (dashed green) for the well-mixed initial distribution with $P/M=1/1000$. (c) $\langle C(t)\rangle$ (solid orange) and $\mathcal{C}(t)$ (dashed green) for the well-mixed initial distribution with $P/M=8/1000$. (d) Vertical strip initial distribution with width $w_1=25$ leading to $\mathcal{C}(0)=0.25$. (e) $\langle C(t)\rangle$ (solid orange) and $\mathcal{C}(t)$ (dashed green) for the vertical strip initial distribution with $P/M=1/1000$. (f) $\langle C(t)\rangle$ (solid orange) and $\mathcal{C}(t)$ (dashed green) for the vertical strip initial distribution with $P/M=8/1000$. (g) Square initial distribution with width $w_1=50$ leading to $\mathcal{C}(0)=0.25$. (h) $\langle C(t)\rangle$ (solid orange) and $\mathcal{C}(t)$ (dashed green) for the square initial distribution with $P/M=1/1000$. (i) $\langle C(t)\rangle$ (solid orange) and $\mathcal{C}(t)$ (dashed green) for the square initial distribution with $P/M=8/1000$. The dashed black horizontal lines in (b)--(c), (e)--(f) and (h)--(i) indicate the Allee threshold, $A=0.4$. }
\label{fig:23} 
\end{figure}

In Figures \ref{fig:5}--\ref{fig:7}, simulations are performed with $M=1$ and $P=1/1000$ meaning that the time scale of migration is 1000 times faster than the time scale of proliferation and death. It is instructive to compare outcomes with $P/M=1/1000$ to those with $P/M=8/1000$ for the well-mixed, vertical strip and square initial distributions in Figure~\ref{fig:23}, where all three initial distributions have $\mathcal{C}(0)=0.25$. Unlike the well-mixed initial distribution, the vertical strip and square initial distributions lead to population survival when $P/M=8/1000$, as shown in Figures~\ref{fig:23}(f) and (i). This is interesting as the global density averaged across the whole domain is smaller than the Allee threshold. This comparison indicates that the vertical strip and square initial distributions may sometimes lead to the survival of the population whereas the same initial number of individuals in a well-mixed environment would lead to extinction. These differences are due to the interplay between the role of the initial spatial distribution and the ratio of time scale of migration to the time scale of proliferation and death.

Overall, the results in Figures~\ref{fig:5}--\ref{fig:23} confirm that the numerical solution of the continuum model provides a useful way of accurately studying the expected behaviour of the discrete model. Of interest is that the long-term fate of populations varies with the spatial arrangement of the initial distributions. Our aim now is to study these differences more carefully.

\section{Role of the shape of the initial distribution}
\label{sec:5}
In this section we first systematically explore the role of the three simple shapes of the initial distribution described in Figure \ref{fig:1}, and then explore the influence of more complicated two-dimensional shapes of the initial distribution on the fate of populations. Our results in Section \ref{sec:4} indicate that several factors are at play when we consider the long-term fate of bistable populations. First, the spatial arrangement of the initial population plays an important role. Second, the ratio $P/M$ also influences the fate of populations. Since the initial distribution of the population is given by Equation \eqref{initialdistribution}, the initial distribution varies with both $B$ and the size of the initially occupied region $\mathcal{H}$ except that the well-mixed initial distribution only varies with $B$. In the remainder of the main document we fix $B=1$ and alter the initial population size by adjusting the size of $\mathcal{H}$ for the initial distributions that are not well-mixed. Furthermore, we always consider an $L\times L$ domain with $L=100$. Additional results in the Supplementary Material explore different choices of $B$ and $L$, and we see that varying these choices does not change our overall observations and conclusions.

\begin{figure}[]
\centering
\includegraphics[width=0.85\textwidth]{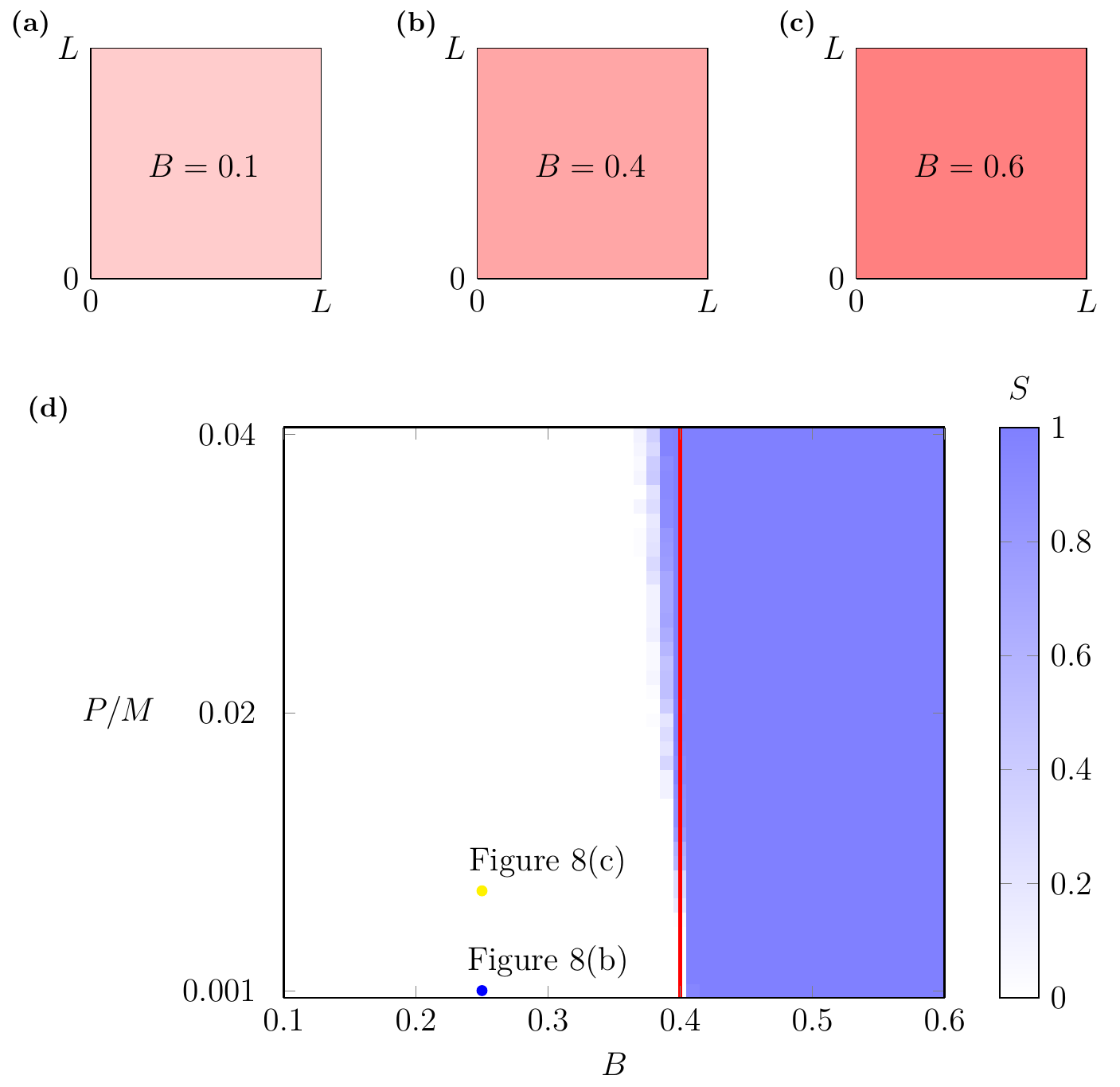}
\singlespace\caption{\textbf{Phase diagram for survival/extinction with the well-mixed initial distribution}. (a)--(c) show how we vary the initial density with $C(0)=B$ for this initial distribution. (d) Phase diagram of a rectangular mesh with $51\times40$ nodes for $C(0)=B\in [1/10, 6/10]$, and $P/M \in [1/1000,4/100]$. The vertical red line indicates the survival/extinction threshold from the continuum model and the blue shading shows the survival probability $S$ measured by 40 identically-prepared realisations. The blue dot indicates the parameters in Figure~\ref{fig:23}(b). The yellow dot indicates the parameters in Figure~\ref{fig:23}(c).}
\label{fig:8} 
\end{figure}

Results in Figure~\ref{fig:8} summarise the long-term outcome of a range of scenarios with the well-mixed initial distribution. In this case $\mathcal{H}$ corresponds to the entire $L \times  L$ domain and $C(0)=B$. We vary the initial distribution by varying $B$, as indicated in Figures~\ref{fig:8}(a)--(c), and vary the ratio $P/M$ by holding $M=1$ and varying $P \in [1/1000, 4/1 00]$ in the discrete model. As $P/M=\lambda/(4D_0)$, we hold $D_0=1/4$ and vary $\lambda$ in the continuum model. To systematically study the transition between population extinction to population survival, we take the $(B,P/M)$ phase space and discretise it uniformly into a rectangular mesh, with $51\times40$ nodes. We note that, unlike the continuum approach that always leads to the same outcome when using the same choice of parameters, different identically-prepared realisations of the stochastic model can lead to different outcomes \citep{anudeep2020,stuart2020predict}. Therefore, for each value of $B$ and $P/M$ considered, we generate 40 identically-prepared realisations of the discrete model and we compute the survival probability, $S \in[0,1]$, as the fraction of realisations in which the population survives after a sufficiently long period of time $\mathcal{T}$, which we take to be $\mathcal{T} = \textrm{max}(30/P,10^4)$. Figure~\ref{fig:8}(d) summarises the outcomes of the simulations in terms of a phase diagram.  In this case the survival outcome for the continuum model is a simple vertical line at $C(0)=A$. In general we see good agreement between the prediction of survival or extinction between the continuum and discrete models.

There are some small discrepancies as $P/M$ increases. In discrete simulations, the local clustering caused by larger $P/M$ leads to higher local densities and thus contributes to the survival of populations, which reflects the influence of stochasticity in the discrete model. This difference is consistent with the fact that for the continuum model $P/M$ has to be sufficiently small, otherwise the mean-field approximation is invalid and the solution of the continuum model does not necessarily provide an accurate description of the discrete mechanism \citep{Baker2010,simpson2010cell}. In summary, for the well-mixed initial distribution the long-term population survival depends simply upon whether the initial density is above or below the Allee threshold, as expected.  

\begin{figure}[]
\centering
\includegraphics[width=0.85\textwidth]{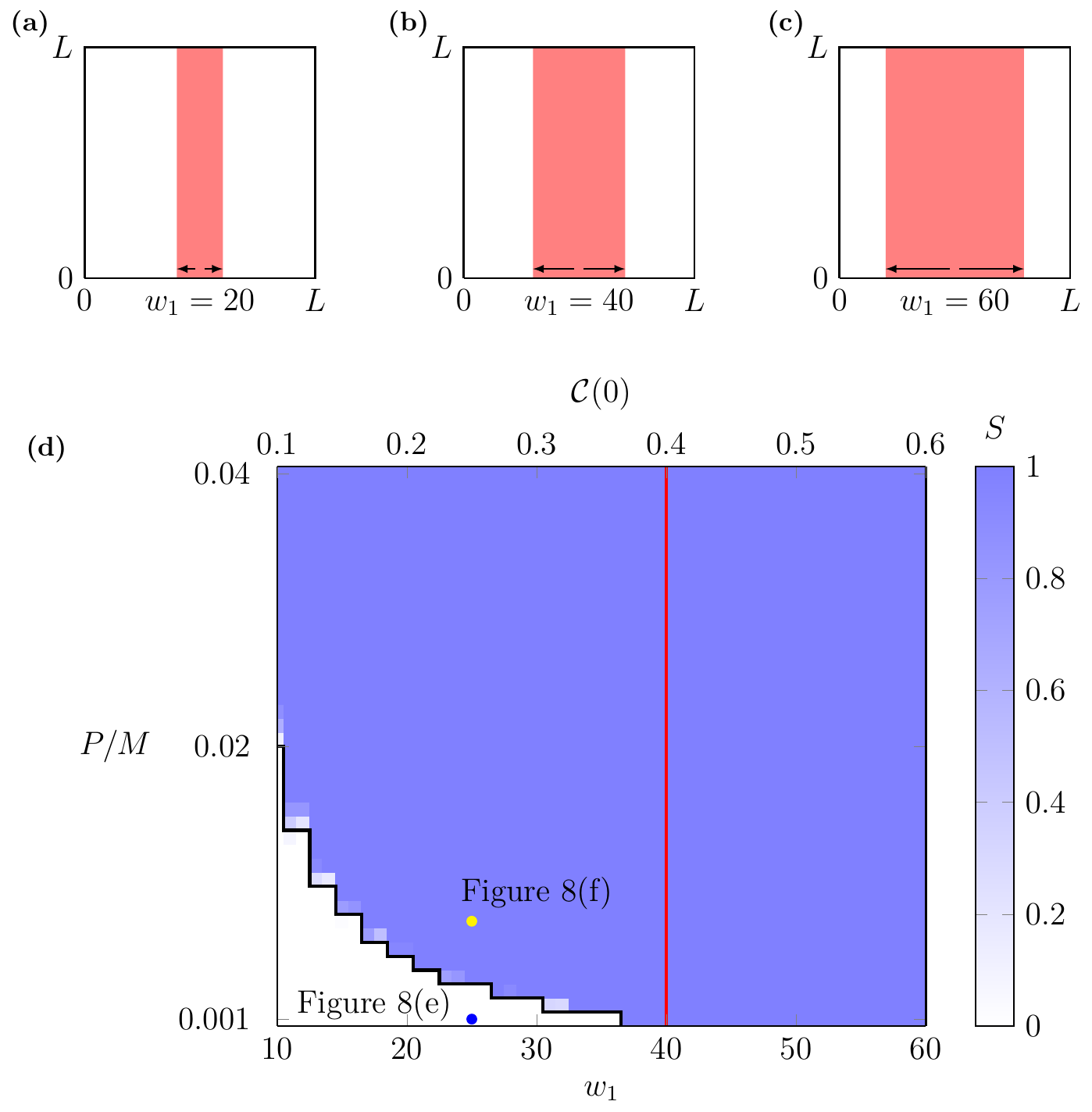}
\singlespace\caption{\textbf{Phase diagram for survival/extinction with the vertical strip initial distribution}. (a)--(c) Three different initial distributions where $\mathcal{C}(0)=w_1/L$, and  we vary $w_1$. (d) Phase diagram of a rectangular mesh with $51\times40$ nodes for $w_1\in[10,60]$, $\mathcal{C}(0) \in [1/10, 6/10]$ and $P/M \in [1/1000,4/100]$. The black curve indicates the survival/extinction threshold from the continuum model and the blue shading shows the survival probability $S$ measured by 40 identically-prepared realisations. The vertical red line is $\mathcal{C}(0)=0.4$ which relates to the Allee threshold, $A=0.4$. The blue dot indicates the parameters in Figure~\ref{fig:23}(e). The yellow dot indicates the parameters in Figure~\ref{fig:23}(f).}
\label{fig:9} 
\end{figure}

We now explore how the simple outcome for the well-mixed initial distribution becomes more complicated when we consider different initial spatial arrangements of the population.  For the vertical strip initial distribution we vary the size of $\mathcal{H}$ by changing the width of the vertical strip, $w_1$. Varying the width of the strip leads to a change in the initial density across the entire domain,  $\mathcal{C}(0)=w_1/L$. For example, Figures~\ref{fig:9}(a)--(c) shows three vertical strip initial distributions with different widths. For these initial distributions we vary the ratio $P/M=\lambda/(4D_0)$ by holding $M=1$ and varying $P$ in the discrete model, and by holding $D_0=1/4$ and varying $\lambda$ in the continuum model. This allows us to consider the $(w_1,P/M)$ phase space, which we discretise into a rectangular mesh with $51\times40$ nodes. Figure~\ref{fig:9}(b) shows a phase diagram illustrating how the survival probability, $S$, depends upon $w_1$ and $P/M$. The boundary that separates the eventual survival and extinction in the continuum model is shown in solid black, and the survival probability from the discrete simulations is shown in blue shading. Overall, the long-term predictions in terms of survival or extinction are consistent between the continuum and discrete models. For completeness we also show the red vertical line indicating the Allee threshold in the sense of global population density averaged across the whole $L\times L$ domain.

It is interesting to compare the results in Figure~\ref{fig:8}(d) and Figure~\ref{fig:9}(d). In the vertical strip case we see that the long-term survival is strongly dependent upon $P/M$ whereas in the well-mixed initial distribution this dependence is less pronounced. Additional results obtained from holding $w_1$ constant and varying $B$ are presented in the Supplementary Material, which show that these two approaches to varying $\mathcal{C}(0)$ lead to very similar outcomes.

\begin{figure}[]
\centering
\includegraphics[width=0.85\textwidth]{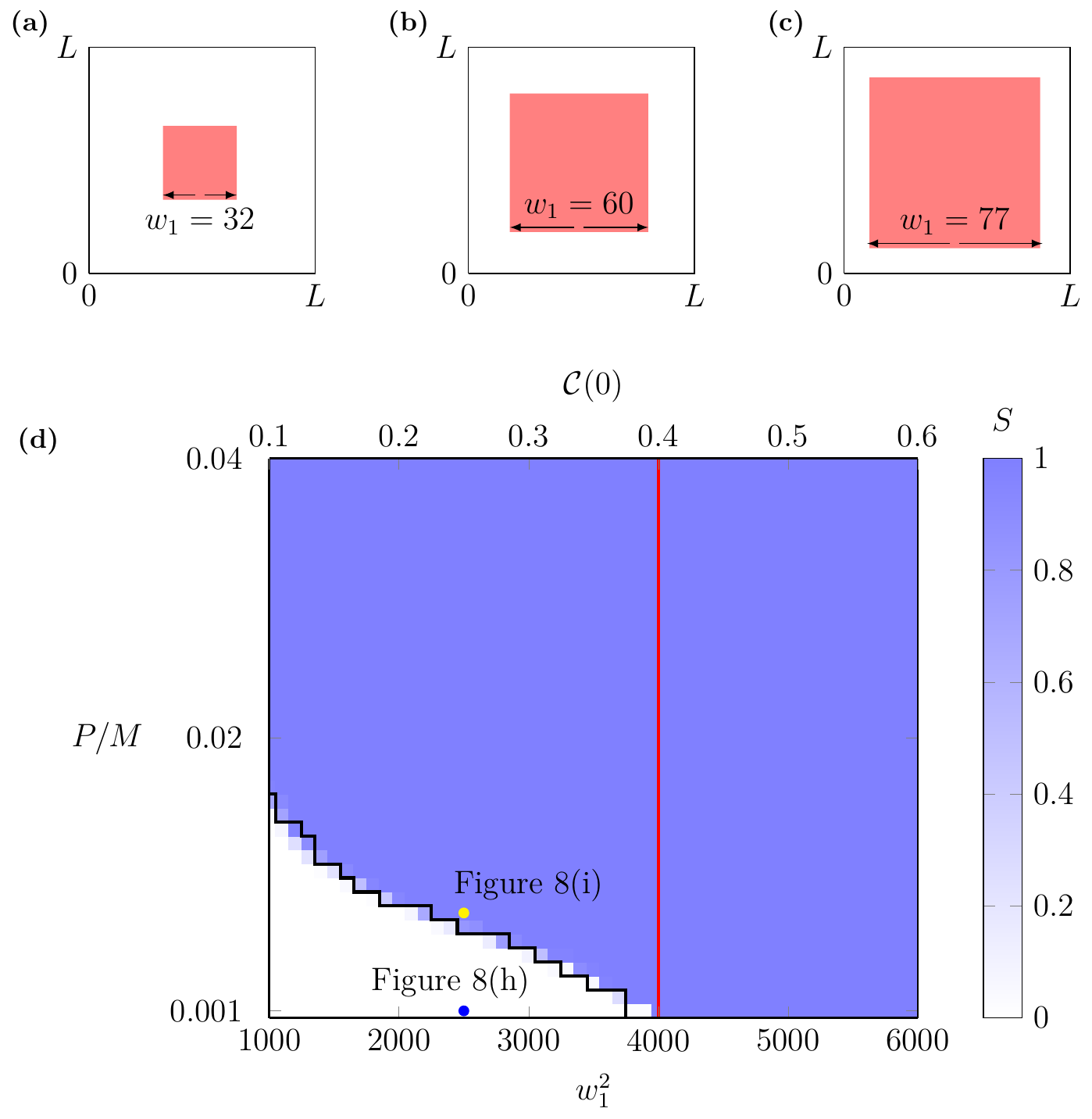}
\singlespace\caption{\textbf{Phase diagram for survival/extinction with the square initial distribution}. (a)--(c) Three different initial distributions where $\mathcal{C}(0)=w_1^2/L^2$, and we vary $w_1$. (d) Phase diagram of a rectangular mesh with $51\times40$ nodes for $w_1^2\in[1000,6000]$, $\mathcal{C}(0) \in [1/10, 6/10]$ and $P/M \in [1/1000,4/100]$ where $M=1$. The black curve indicates the survival/extinction threshold from the continuum model and the blue shading shows the survival probability $S$ measured by 40 identically-prepared realisations. The vertical red line is $\mathcal{C}(0)=0.4$ which relates to the Allee threshold, $A=0.4$. The blue dot indicates the parameters in Figure~\ref{fig:23}(h). The yellow dot indicates the parameters in Figure~\ref{fig:23}(i).}
\label{fig:10} 
\end{figure}

\newpage
For the square initial distribution, we vary the size of $\mathcal{H}$ by changing $w_1$ as shown in Figures~\ref{fig:10}(a)--(c). Varying $w_1$ allows us to vary the initial density across the entire domain, $\mathcal{C}(0)=w_1^2/L^2$. Similar to Figure~\ref{fig:9}, we construct a phase diagram in Figure~\ref{fig:10}(d) that summarises the long-term survival outcome as a function of $w_1^2$ and $P/M$, by discretising the $(w_1^2, P/M)$ phase space using a rectangular mesh with $51\times40$ nodes. The phase diagram in Figure~\ref{fig:10}(d) is very similar to the phase diagram in Figure~\ref{fig:9}(d). We see that the long-term survival strongly depends upon $P/M$, and the distinction between survival and extinction predicted by the continuum limit model is a good approximation of the discrete simulation data. Again, we show additional results obtained from holding $w_1$ constant and varying $B$ in the Supplementary Material, which show that these two approaches to varying $\mathcal{C}(0)$ lead to very similar outcomes.

\newpage
Results in Figures~\ref{fig:8}--\ref{fig:10} show that the long-term survival of a population depends upon $P/M$ and the initial arrangement of the population in a complicated manner. Stochasticity only plays a role on the fate of populations in the discrete model when parameters are close to the boundary that separates the eventual survival and extinction in the continuum model. A key feature of the initial shape is the dimension of the shape. The well-mixed, vertical strip and square initial distributions can be thought of as zero-, one- and two-dimensional shapes, respectively. To highlight the influence of the dimensionality on the fate of the population, we consider a rectangular
distribution of varying initial heights, see Figures~\ref{fig:11}(g)--(i). The initially occupied region $\mathcal{H}$ is a rectangle with width $w_1=40$ and height $w_2\in [25,100]$, which leads to $\mathcal{C}(0)\in[0.1,0.4]$. When $\mathcal{C}(0)=0.16$ with $w_2=40$, the rectangular initial distribution is the same as the square initial distribution, as shown in Figures~\ref{fig:11}(e) and (h). When $\mathcal{C}(0)=0.4$ with $w_2=100$, the rectangular initial distribution is the same as the vertical strip initial distribution, as shown in Figures~\ref{fig:11}(c) and (i). Note that we use $\mathcal{C}(0)$ as the horizontal axis in the phase diagram so that we can compare the results with different shapes of initial distributions. We show the evolution of the total population density in the continuum model with $M=1$ and $P=0.0028$, leading to $D_0=1/4$ and $\lambda=0.0028$, and different $\mathcal{C}(0)$ in Figures~\ref{fig:11}(j)--(m). When $\mathcal{C}(0)=0.2$ and $\mathcal{C}(0)=0.3$, the rectangular initial distribution leads to extinction, which is the same as the results obtained from the square initial distribution. While the vertical strip initial distribution leads to survival with $\mathcal{C}(0)=0.3$.  In contrast, when $\mathcal{C}(0)=0.33$ and $\mathcal{C}(0)=0.36$, the rectangular initial distribution leads to survival, which is the same as the results obtained from the vertical strip initial distribution. While the square initial distribution leads to extinction with $\mathcal{C}(0)=0.33$. This indicates a switch of the influence of the rectangular initial distribution on the fate of populations from a manner similar to the square initial distribution to a manner similar to the vertical strip initial distribution. In Figure~\ref{fig:11}(n), we draw the survival/extinction boundary from the continuum model with the rectangular initial distribution in the $(\mathcal{C}(0),P/M)$ phase space for $\mathcal{C}(0)\in[0.1,0.4]$ and $P/M=[1/10000,21/1000]$, and compare them to the results obtained from the vertical strip initial distribution in Figure~\ref{fig:9} and from the square initial distribution in Figure~\ref{fig:10}. Although the results are from the continuum model, we still use $P/M$ as the vertical axis to reflect the connection between the discrete and continuum models in our framework. We observe that there is a clear transition of the survival/extinction boundary for the rectangular initial distribution. The survival/extinction boundary of the rectangular initial distribution is close to the survival/extinction boundary obtained from the square initial distribution when $\mathcal{C}(0)$ is small, and 
is close to the survival/extinction boundary obtained from the vertical strip initial distribution when $\mathcal{C}(0)$ is large. This transition indicates that the dimensionality of the initial shape of a population plays a role in determining the ultimate fate of the population.

\afterpage{
\thispagestyle{empty}
\begin{figure}[]
\vspace{-3cm}
\centering
\includegraphics[width=0.8\textwidth]{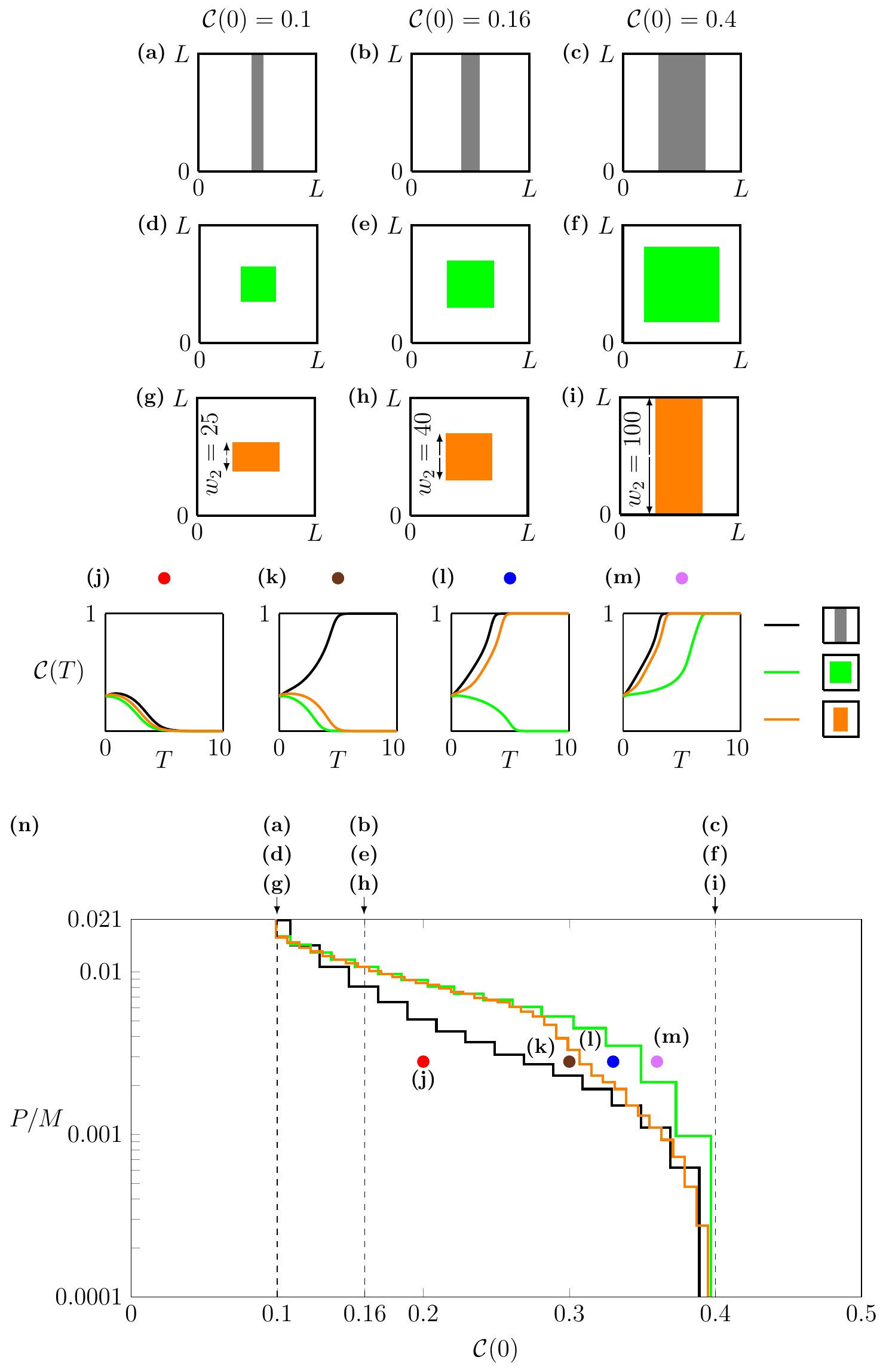}
\singlespace\caption{
\textbf{Influence of dimensionality in rectangular initial distributions.} (a)--(c) Vertical strip (one-dimensional) initial distributions varied with the width $w_1\in[10,40]$. (d)--(f) Square (two-dimensional) initial distributions varied with the width $w_1\in[32,64]$. (g)--(i) The rectangular initial distributions varied with the height $w_2\in[25,100]$ with a fixed width $w_1=40$. (j)--(m) The evolution of the total population density $\mathcal{C}(T)$ with $P/M=0.0028$, $\mathcal{C}(0)=0.2$ in (j), 0.3 in (k), 0.33 in (l) and 0.36 in (m), and with the vertical strip (black), square (green) and rectangular (orange) initial distributions. (n) Phase diagram showing the survival/extinction boundaries constructed from a $151\times120$ array of $\mathcal{C}(0)\in[1/10,1/40]$ and $P/M\in[1/10000,21/1000]$. Three curves indicate the survival/extinction thresholds from the continuum model of the vertical strip (black), square (green) and rectangular (orange) initial distributions. Three black dashed lines represent $\mathcal{C}(0)=0.1, 0.16$ and $0.4$. Note that we use a logarithmic scale for the $P/M$ axis.}
\label{fig:11}
{\makebox[\linewidth]{\thepage}}
\end{figure}
\clearpage
}

\newpage

\begin{figure}[]
\centering
\includegraphics[width=\textwidth]{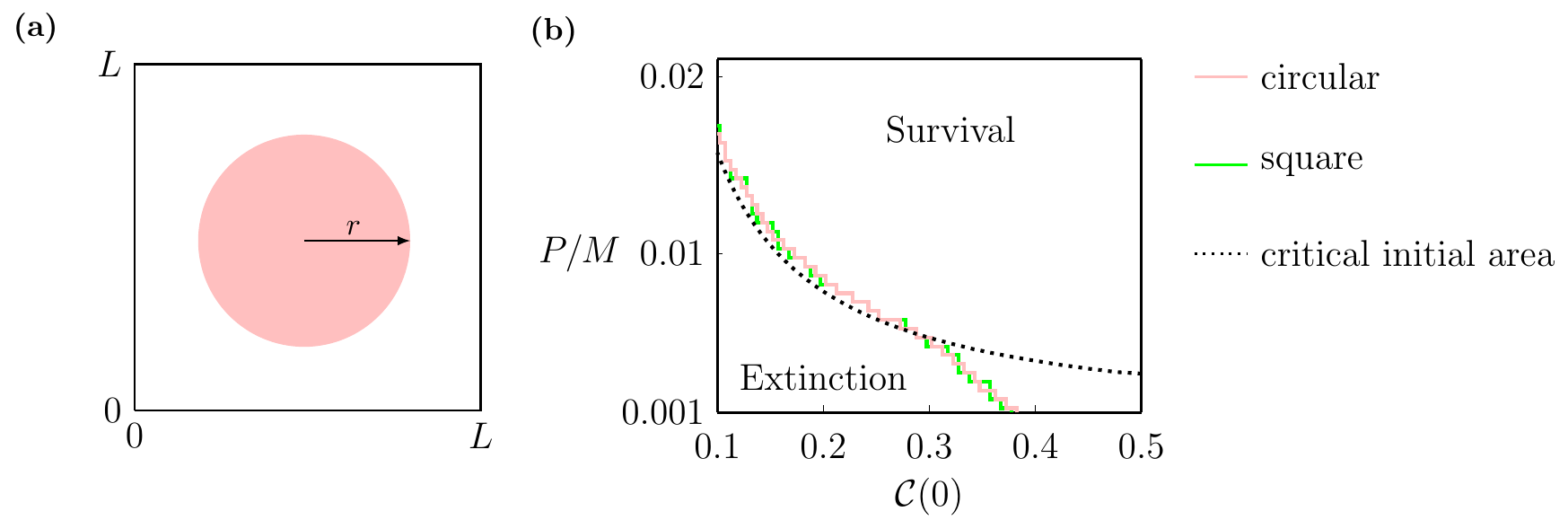}
\singlespace\caption{
\textbf{Phase diagram for survival/extinction with the circular initial distributions.} (a) A circular initial distribution with radius $r$. (b) Phase diagram on a rectangular mesh with $61\times41$ nodes for $\mathcal{C}(0) \in [0.1, 0.5]$ and $P/M \in [1/1000,21/1000]$ where $M=1$. Pink curve indicates the survival/extinction threshold in the continuum model with the circular initial distributions. Black dotted curve indicates the survival/extinction threshold obtained from \eqref{area}. Green curve indicates the survival/extinction threshold for the square initial distributions, as shown in Figure~\ref{fig:10}.}
\label{fig:12} 
\end{figure}

Many more spatial arrangements of the population can be considered. We first consider a circle with radius $r$ as the initially occupied region $\mathcal{H}$, as shown in Figure~\ref{fig:12}(a). We draw the phase diagram from the continuum model by varying $P/M\in[1/1000,21/1000]$ where $M=1$ and $\mathcal{C}(0)=\pi r^2/L^2\in[0.1,0.5]$ with $r\in[17.8,39.9]$ in Figure~\ref{fig:12}(b). We then consider the critical initial radius 
\begin{equation}
    \label{radius}
    r_{\text{crit}}=\sqrt{\frac{2D_0}{\lambda a}}\frac{1}{1-2A},
\end{equation}
derived by \cite{lewis1993allee}, which leads to the critical initial area $A_{\text{crit}}=\pi r_{\text{crit}}^2$. As $\lambda$ and $D_0$ depend on $P$ and $M$ in our framework, we derive the survival/extinction threshold of the initial total population density 
\begin{equation}
    \label{area}
    \mathcal{C}(0)_{\text{crit}}=\frac{\pi M}{2aP}\frac{1}{(1-2A)^2L^2},
\end{equation}
and draw the extinction/survival
boundary in the $(\mathcal{C}(0),P/M)$ phase space based on \eqref{area} in Figure~\ref{fig:12}(b). Although the critical initial radius is formally derived in the limit $P/M\gg1$, this result also appears to work well here where $P/M$ is not that large. Furthermore, we compare the survival/extinction boundary to the result obtained from the square initial distribution in Figure~\ref{fig:12}(b). These two survival/extinction boundaries are very close, which indicates that the square initial distribution and the circular initial distribution give rise to similar outcomes. This could be attributed to the fact that they are both compact
initial distributions with a small perimeter to area ratio.

\begin{figure}[]
\centering
\includegraphics[width=\textwidth]{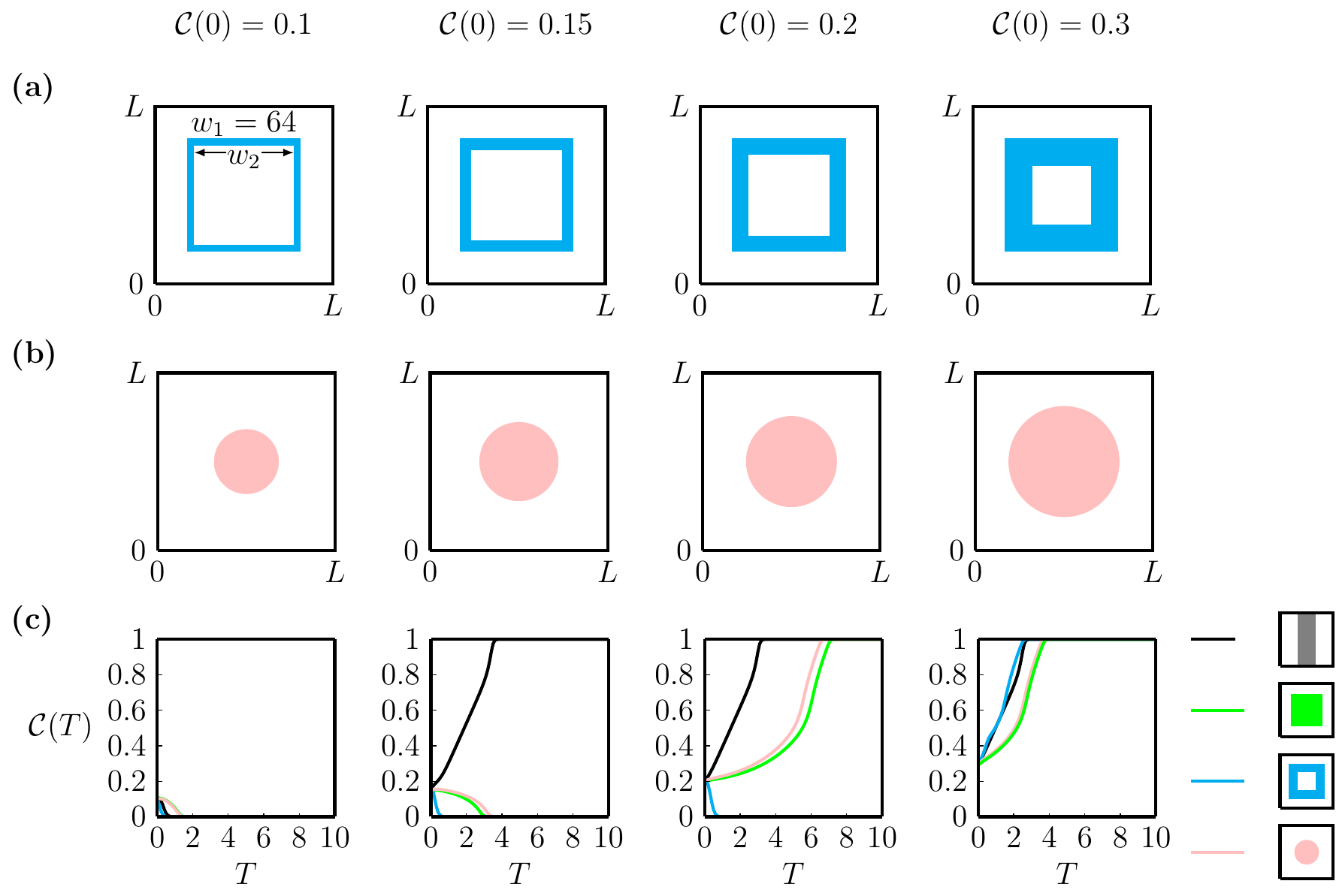}
\singlespace\caption{\textbf{Population dynamics with more complicated initial spatial distributions}. (a) Square annular initial distributions with a fixed outer width $w_1=64$ and different values of inner width $w_2$. (b) Circular initial distributions with different values of radius $r$. (c) The evolution of the total population density $\mathcal{C}(T)$ which considers $P/M=0.01$, with $P=0.01$ and $M=1$, and $\mathcal{C}(0)=0.1,0.15,0.2,0.3$ with different initial distributions. Black curves are generated by the vertical strip initial distributions. Green curves are generated by the square initial distributions. Cyan curves are generated by the square annular initial distributions. Pink curves are generated by the circular initial distributions.}
\label{fig:13} 
\end{figure}

A natural question is whether populations with other two-dimensional initial distributions have the similar critical initial area determined by \eqref{radius}. To explore this, we now consider a square annulus in the middle of the domain as the initially occupied region, as shown in Figure~\ref{fig:13}(a). The area of the region is determined by a fixed outer width $w_1=64$ and a variable inner width $w_2$. We also consider a circle as the initially occupied region, as shown in Figure~\ref{fig:13}(b). The area of the region varies with radius $r$. We show the evolution of the total population density in the continuum model with these two initial shapes at $P/M=0.01$, where $M=1$ and $P=0.01$ leading to $D_0=1/4$ and $\lambda=0.01$, and different $\mathcal{C}(0)$. We further show the results obtained from the vertical strip initial distributions and the square initial distributions in Figure~\ref{fig:13}(c). All four initial distributions lead to population extinction when $\mathcal{C}(0)=0.1$. When $\mathcal{C}(0)=0.15$, only the vertical strip initial distribution leads to population survival. When $\mathcal{C}(0)=0.2$, the square and circular initial distributions also lead to population survival. In contrast, the square annular initial distribution still leads to population extinction. Note that $P/M=0.01$ leads to $A_\text{crit}\approx1570$ and $\mathcal{C}(0)_{\text{crit}}=0.157<0.2$. This suggests that, although the area of the square annulus exceeds the critical initial area given in \cite{lewis1993allee}, it still leads to population extinction. When $\mathcal{C}(0)=0.3$, all four initial distributions lead to population survival. These results indicate that, although some initial distributions have the same area of the initially occupied region, they may lead to different fates of the population. It is the shape of the initially occupied region that dictates whether a bistable population survives or goes extinct. This suggests the importance of considering the influence of the spatial arrangements of individuals on the long-term survival of populations.

\section{Conclusions and Outlook}
\label{sec:7}
In this work we design, analyse and implement a new two-dimensional stochastic discrete model incorporating movement, birth and death events with crowding effects to study population extinction. The continuum limit of the discrete model is a nonlinear RDE which can be used to study a wide range of macroscopic phenomena including linear diffusion, nonlinear diffusion, as well as logistic and bistable growth kinetics.  Since the aim of this work is to focus on long term survival or extinction, we choose the movement crowding function to be $G(C)=1-C$ which corresponds to macroscopic linear diffusion. In addition, we choose the growth crowding function to be $F(C) = a(1-C)(C-A)$ which leads to a classical cubic bistable source term with Allee threshold~$A$.  Using a range of initial conditions, we show that numerical solutions of the continuum RDE compare well with appropriately averaged data from the discrete model.

The focus of our work is to use the discrete and continuum models to explore the factors that influence the long-term fate of the bistable population. In particular, we explore different spatial arrangements of the population on a finite $L \times L$ domain with periodic boundary conditions.  The well-mixed initial distribution involves distributing agents evenly across the entire $L \times L$ domain, the vertical strip initial distribution involves distributing agents along a vertical strip within the $L \times L$ domain so that the initial density is independent of vertical position in the domain, and the two-dimensional initial distributions involve distributing agents in a square, circular, rectangular or square annular region within the $L \times L$ domain. Our results show that the shape of initial distributions plays an important role in determining the fate of populations. This suggests the importance of considering the influence of spatial arrangements of individuals in studies of population dynamics.

There are many avenues for extending the work presented in this study. The stochastic model provides very detailed information including the age structure of the population and individual trajectories, see the results in the Supplementary Material. Furthermore, Other shapes of initial distributions than those investigated here can be considered and similar numerical explorations of the long-term survival or extinction of the populations can be conducted using the software provided on \href{https://github.com/oneflyli/Yifei2020Dimensionality}{GitHub} for both the continuum and discrete models. Another feature of this work that could be explored is the choice of crowding functions.  As we pointed out, all simulations here focus on $G(C)=1-C$, which gives rise to linear diffusion, and $F(C) = a(1-C)(C-A)$ which gives rise to the classical cubic bistable term. Other choices of $G(C)$ and $F(C)$ can be incorporated into the discrete model to explore how the results presented here depend upon the precise details of these choices of crowding functions.  We note that other choices of $G(C)$ lead to different motility mechanisms that are associated with nonlinear diffusion mechanisms, and that these can be important for applications where adhesion~\citep{Deroulers2009} and inertial effects~\citep{Stephen2019} are relevant.  While we have not explicitly explored these effects in this work, our framework is sufficiently general that these mechanisms can be incorporated and explored, if required. Moreover, other boundary conditions could be incorporated in our model. In the Supplementary Material, we show that no-flux boundary conditions lead to the same result as when we consider periodic boundary conditions. Another interesting extension would be to consider Allee-type dynamics with populations of interacting species~\citep{Mat2009MultiExclusion}. Under these conditions interactions can also contribute to the eventual survival or extinction of any of the subpopulations~\citep{Taylor2020,Krause2020}.

\noindent
\textit{Acknowledgements:} This work is supported by the Australian Research Council (DP200100177, DE200100988, DP190102545). We thank the two referees and the handling editor for their helpful suggestions.

\appendix 

\renewcommand{\thealgocf}{S\arabic{algocf}}

\section{Algorithm for discrete simulations}\setstretch{0.8}
\begin{tcolorbox}
\begin{algorithm}[H]
\SetAlgoLined
 Create a two-dimensional $I\times J$ hexagonal lattice; Distribute agents with specific initial conditions; The total number of lattice site is $IJ$\;
 Set $t=0$; Calculate total agents $Q(t)$\;
 \While{\normalfont $t<t_{\text{end}}$ and $Q(t)>0$ and $Q(t)\le IJ$  }{
 $t=t+\tau$\;
 $Q(t)=Q(t-\tau)$\;
 $B_1=0$; $B_2=0$\;
 Draw two random variables: $\beta_1\sim \textit{U}[0,1]$, $\beta_2\sim \textit{U}[0,1]$\;
 \While{$B_1<Q(t)$ }{
 $B_1=B_1+1$\;
 Randomly choose an agent $\mathbf{s}$\;
  \uIf{$\beta_1<M$}{
   Calculate $\newbar{K}_{\mathbf{s}}^{(\textls{m})}$ and $G(\newbar{K}_{\mathbf{s}}^{(\textls{m})})$\;
   Draw a random variable: $\gamma_1\sim \textit{U}[0,1]$\;
    \uIf{\normalfont $\gamma_1<G(\newbar{K}_{\mathbf{s}}^{(\text{m})})$}{
       Randomly choose a vacant site in $\mathcal{N}_{1}(\mathbf{s})$ and move agent to chosen site
       }
    \Else{
   Nothing happens\;
  }
  }
  \Else{
   Nothing happens\;
  }
 }
  \While{$B_2<N(t)$ }{
 $B_2=B_2+1$\;
 Randomly choose an agent $\mathbf{s}$\;
  \uIf{$\beta_2<P$}{
   Calculate ${K}_{\mathbf{s}}^{(\textls{g})}$ and $F(\newbar{K}_{\mathbf{s}}^{(\textls{g})})$\;
   Calculate a random variable: $\gamma_2\sim \textit{U}[0,1]$\;
   \uIf{\normalfont $F(\newbar{K}_{\mathbf{s}}^{(\text{g})})>0$}{
       \uIf{\normalfont $\gamma_2<F(\newbar{K}_{\mathbf{s}}^{(\text{g})})$}{
       Randomly choose a vacant site in $\mathcal{N}_{4}(\mathbf{s})$ and place a new agent on chosen site\;
       $Q(t)=Q(t)+1$
       }
   }
   \uElseIf{\normalfont $F(\newbar{K}_{\mathbf{s}}^{(\text{g})})<0$}{
    \uIf{\normalfont $\gamma_2<-F(\newbar{K}_{\mathbf{s}}^{(\text{g})})$}{
       Remove agent\;
       $Q(t)=Q(t)-1$\;
       }
  }
    \Else{
   Nothing happens\;
  }
  }
  \Else{
   Nothing happens\;
  }
 }
 }
\caption{Pseudo-code for a single realisation of the stochastic model}
\end{algorithm}
\end{tcolorbox}

\setstretch{1.5}
\section{Derivation of the continuum limit}
We recall Equation~(5), that is, the expected change in occupancy of site $\mathbf{s}$ during the time interval from $t$ to $t+\tau$,
\begin{equation}
\label{delta1_s}
     \begin{aligned}
     \delta(\newbar{C}_{\mathbf{s}})=&\frac{M}{\lvert\mathcal{N}_{1}\rvert}(1-\newbar{C}_{\mathbf{s}})\sum_{\mathbf{s}'\in \mathcal{N}_{1}\{\mathbf{s}\}}\newbar{C}_{\mathbf{s}'}\frac{G(\newbar{K}_{\mathbf{s}'}^{(\textls{m})})}{1-\newbar{K}_{\mathbf{s}'}^{(\textls{m})}}-M\newbar{C}_{\mathbf{s}}G(\newbar{K}_{\mathbf{s}}^{(\textls{m})})\\
          &+\frac{P}{\lvert\mathcal{N}_{4}\rvert}(1-\newbar{C}_{\mathbf{s}})\sum_{\mathbf{s}'\in \mathcal{N}_{4}\{\mathbf{s}\}}\mathbbm{H}(F(\newbar{K}_{\mathbf{s}'}^{(\textls{g})}))\newbar{C}_{\mathbf{s}'}\frac{F(\newbar{K}_{\mathbf{s}'}^{(\textls{g})})}{1-\newbar{K}_{\mathbf{s}'}^{(\textls{g})}}-(1-\mathbbm{H}(F(\newbar{K}_{\mathbf{s}}^{(\textls{g})}))P\newbar{C}_{\mathbf{s}}F(\newbar{K}_{\mathbf{s}}^{(\textls{g})}).
     \end{aligned}
\end{equation}
As we know that the continuum limit of the last two terms in Equation \eqref{delta1_s} leads to a source term $\lambda C F(C)$ \citep{jin2016stochastic}, we focus on the movement mechanism, that is, the first two terms on the right hand side of Equation~\eqref{delta1_s}. For convenience, we will omit the overlines on notations in the following content. 

It is useful to first write the general form of the Taylor series relating the occupancy of sites $(x+a,y+b)$,
\begin{equation}
    \label{general}
    C_{x+a,y+b}=C_{x,y}+\frac{(a\Delta)^1}{1!}\frac{\partial C_{x,y}}{\partial x}+\frac{(b\Delta)^1}{1!}\frac{\partial C_{x,y}}{\partial y}+\frac{(a\Delta)^2}{2!}\frac{\partial C^2_{x,y}}{\partial x^2}+\frac{2ab\Delta^2}{2!}\frac{\partial C^2_{x,y}}{\partial x\partial y}+\frac{(b\Delta)^2}{2!}\frac{\partial C^2_{x,y}}{\partial y^2}+\mathcal{O}(\Delta^3).
\end{equation}
We represent the six nearest neighbouring sites of site~$\mathbf{s}$ located at $(x,y)$ as site $\mathbf{s}_1$ with $(x-\Delta,y)$; site $\mathbf{s}_2$ with $(x+\Delta,y)$; site $\mathbf{s}_3$ with $(x-\Delta/2,y+\Delta\sqrt{3}/2)$; site $\mathbf{s}_4$ with $(x+\Delta/2,y+\Delta\sqrt{3}/2)$; site $\mathbf{s}_5$ with $(x-\Delta/2,y-\Delta\sqrt{3}/2)$ and site $\mathbf{s}_6$ with $(x+\Delta/2,y-\Delta\sqrt{3}/2)$. That is, $\mathcal{N}_1=\{\mathbf{s}_1,\mathbf{s}_2,\mathbf{s}_3,\mathbf{s}_4,\mathbf{s}_5,\mathbf{s}_6\}$. The truncated Taylor series of these sites are
\begin{align}
    &C_{\mathbf{s}_1}=C_{\mathbf{s}}-\frac{\partial C_{\mathbf{s}}}{\partial x}\Delta+\frac{\partial^2 C_{\mathbf{s}}}{\partial x^2}\frac{\Delta^2}{2}+\mathcal{O}(\Delta^3),\label{Tayler1}\\
    &C_{\mathbf{s}_2}=C_{\mathbf{s}}+\frac{\partial C_{\mathbf{s}}}{\partial x}\Delta+\frac{\partial^2 C_{\mathbf{s}}}{\partial x^2}\frac{\Delta^2}{2}+\mathcal{O}(\Delta^3),\label{Taylor2}\\
    &C_{\mathbf{s}_3}=C_{\mathbf{s}}-\frac{\partial C_{\mathbf{s}}}{\partial x}\frac{\Delta}{2}+\frac{\partial C_{\mathbf{s}}}{\partial y}\frac{\sqrt{3}\Delta}{2}+\left[\frac{1}{4}\frac{\partial^2 C_{\mathbf{s}}}{\partial x^2}+\frac{3}{4}\frac{\partial^2 C_{\mathbf{s}}}{\partial y^2}-\frac{\sqrt{3}}{2}\frac{\partial^2 C_{\mathbf{s}}}{\partial x\partial y}\right]\frac{\Delta^2}{2}+\mathcal{O}(\Delta^3),\label{Tayler3}\\
    &C_{\mathbf{s}_4}=C_{\mathbf{s}}+\frac{\partial C_{\mathbf{s}}}{\partial x}\frac{\Delta}{2}+\frac{\partial C_{\mathbf{s}}}{\partial y}\frac{\sqrt{3}\Delta}{2}+\left[\frac{1}{4}\frac{\partial^2 C_{\mathbf{s}}}{\partial x^2}+\frac{3}{4}\frac{\partial^2 C_{\mathbf{s}}}{\partial y^2}+\frac{\sqrt{3}}{2}\frac{\partial^2 C_{\mathbf{s}}}{\partial x\partial y}\right]\frac{\Delta^2}{2}+\mathcal{O}(\Delta^3),\label{Tayler4}\\
    &C_{\mathbf{s}_5}=C_{\mathbf{s}}-\frac{\partial C_{\mathbf{s}}}{\partial x}\frac{\Delta}{2}-\frac{\partial C_{\mathbf{s}}}{\partial y}\frac{\sqrt{3}\Delta}{2}+\left[\frac{1}{4}\frac{\partial^2 C_{\mathbf{s}}}{\partial x^2}+\frac{3}{4}\frac{\partial^2 C_{\mathbf{s}}}{\partial y^2}+\frac{\sqrt{3}}{2}\frac{\partial^2 C_{\mathbf{s}}}{\partial x\partial y}\right]\frac{\Delta^2}{2}+\mathcal{O}(\Delta^3),\label{Tayler5}\\
    &C_{\mathbf{s}_6}=C_{\mathbf{s}}+\frac{\partial C_{\mathbf{s}}}{\partial x}\frac{\Delta}{2}-\frac{\partial C_{\mathbf{s}}}{\partial y}\frac{\sqrt{3}\Delta}{2}+\left[\frac{1}{4}\frac{\partial^2 C_{\mathbf{s}}}{\partial x^2}+\frac{3}{4}\frac{\partial^2 C_{\mathbf{s}}}{\partial y^2}-\frac{\sqrt{3}}{2}\frac{\partial^2 C_{\mathbf{s}}}{\partial x\partial y}\right]\frac{\Delta^2}{2}+\mathcal{O}(\Delta^3).\label{Tayler6}
\end{align}
The local density of $\mathbf{s}$ is obtained by summing the Taylor series of sites in $\mathcal{N}_1\{\mathbf{s}\}$, that is,
\begin{equation}
\label{averages_c}
\begin{aligned}
    K_{\mathbf{s}}^{(\textls{m})}&=\frac{1}{6}\sum_{\mathbf{s}''\in\mathcal{N}_1\{\mathbf{s}\}}C_{\mathbf{s}''}\\
    &=C_{\mathbf{s}}+\left(\frac{\partial^2C_{\mathbf{s}}}{\partial x^2}+\frac{\partial^2C_{\mathbf{s}}}{\partial y^2}\right)\frac{\Delta^2}{4}+\mathcal{O}(\Delta^3).
\end{aligned}
\end{equation}
Similarly, the local density of $\mathbf{s_1}$ is obtained by summing the Taylor series of sites in $\mathcal{N}_1\{\mathbf{s}_1\}$, that is,
\begin{equation}
\label{averages'}
\begin{aligned}
    K_{\mathbf{s}_1}^{(\textls{m})}&=\frac{1}{6}\sum_{\mathbf{s}''\in\mathcal{N}_1\{\mathbf{s}_1\}}C_{\mathbf{s}''}\\
    &=C_{\mathbf{s}_1}+\left(\frac{\partial^2C_{\mathbf{s}_1}}{\partial x^2}+\frac{\partial^2C_{\mathbf{s}_1}}{\partial y^2}\right)\frac{\Delta^2}{4}+\mathcal{O}(\Delta^3),\\
    &=C_{\mathbf{s}}-\frac{\partial C_{\mathbf{s}}}{\partial x}\Delta+\frac{\partial^2C_{\mathbf{s}}}{\partial x^2}\frac{\Delta^2}{2}+\left(\frac{\partial^2C_{\mathbf{s}}}{\partial x^2}+\frac{\partial^2C_{\mathbf{s}}}{\partial y^2}\right)\frac{\Delta^2}{4}+\mathcal{O}(\Delta^3).
\end{aligned}
\end{equation}
For simplification we rewrite Equation \eqref{averages'} as $K_{\mathbf{s}_1}^{(\textls{m})}=C_\mathbf{s}+\widetilde{C}_{\mathbf{s}_1}$, where $\widetilde{C}_{\mathbf{s}_1}\sim\mathcal{O}(\Delta)$. Subsequently, the movement crowding function at $\mathbf{s}_1$ can be expanded as
\begin{equation}
    \label{crowding1}
    \begin{aligned}
        G\left(K_{\mathbf{s}_1}^{(\textls{m})}\right)&=G\left(C_\mathbf{s}+\widetilde{C}_{\mathbf{s}_1}\right),\\
        &=G\left(C_\mathbf{s}\right)+\frac{\textrm{d}G\left(C_\mathbf{s}\right)}{\textrm{d}C}\widetilde{C}_{\mathbf{s}_1}+\frac{\textrm{d}^2G\left(C_\mathbf{s}\right)}{\textrm{d}C^2}\frac{{\widetilde{C}_{\mathbf{s}_1}}^2}{2}.
    \end{aligned}
\end{equation}
The expansions of $G(K_{\mathbf{s}_2}^{(\textls{m})})$, $G(K_{\mathbf{s}_3}^{(\textls{m})})$,...,$G(K_{\mathbf{s}_6}^{(\textls{m})})$ have similar forms to \eqref{crowding1}. We then go back to the first term on the right hand side of \eqref{delta1_s}, which gives
\begin{equation}
\label{first_1}
        \frac{M}{6}(1-C_\mathbf{s})\sum_{\mathbf{s}'\in \mathcal{N}_1\{\mathbf{s}\}}C_{\mathbf{s}'}\frac{G(K_{\mathbf{s}'}^{(\textls{m})})}{1-K_{\mathbf{s}'}^{(\textls{m})}}.
\end{equation}
For convenience we further drop the $\mathbf{s}$ notation so that $C_\mathbf{s}$ becomes $C$ and $C_{\mathbf{s}_1}$ becomes $C_{1}$. Subsequently, \eqref{first_1} becomes
\begin{equation}
\label{first_2}
        \frac{M}{6}(1-C)\sum_{i=1}^6C_i\frac{G(K_{\mathbf{s}_i}^{(\textls{m})})}{1-K_{\mathbf{s}_i}^{(\textls{m})}}.
\end{equation}
Moreover, we will use two notations
\begin{equation}
    \mathcal{A}=\left(\frac{\partial^2C_{\mathbf{s}}}{\partial x^2}+\frac{\partial^2C_{\mathbf{s}}}{\partial y^2}\right)\frac{\Delta^2}{4}, \quad 
    \mathcal{B}=\left(\left(\frac{\partial C_{\mathbf{s}}}{\partial x}\right)^2+\left(\frac{\partial C_{\mathbf{s}}}{\partial y}\right)^2\right)\frac{\Delta^2}{4},
\end{equation}
in the following content. Expanding the term related to site $\mathbf{s}_1$ in \eqref{first_2} gives
\begin{equation}
    \nonumber
    \begin{aligned}
    &\dfrac{M}{6}(1-C)\left(C+\widetilde{C}_1-\mathcal{A}\right)\frac{\left(G(C)+G'(C)\widetilde{C}_1+G''(C)\dfrac{\widetilde{C}^2_1}{2}\right)}{1-\left(C+\widetilde{C}_1\right)}\\
    =&\dfrac{M}{6}(1-C)\left(C+\widetilde{C}_1-\mathcal{A}\right){\left(G(C)+G'(C)\widetilde{C}_1+G''(C)\dfrac{\widetilde{C}_1^2}{2}\right)}\left(\frac{1}{1-C}+\frac{{\widetilde{C}_1}}{(1-C)^2}+\frac{{\widetilde{C}_1}^2}{(1-C)^3}\right)+\mathcal{O}(\Delta^3)\\
    =&\dfrac{M}{6}\left(C+\widetilde{C}_1-\mathcal{A}\right){\left(G(C)+G'(C)\widetilde{C}_1+G''(C)\dfrac{\widetilde{C}_1^2}{2}\right)}\left(1+\frac{{\widetilde{C}_1}}{1-C}+\frac{{\widetilde{C}_1}^2}{(1-C)^2}\right)+\mathcal{O}(\Delta^3)\\
    =&\dfrac{M}{6}\left[CG(C)+\left(CG'(C)+\frac{G(C)}{1-C}\right)\widetilde{C}_1+\left(\frac{G(C)}{(1-C)^2}+\dfrac{G'(C)}{1-C}+\frac{CG''(C)}{2}\right){\widetilde{C}_1^2}-{G(C)}\mathcal{A}\right]+\mathcal{O}(\Delta^3).
    \end{aligned}
\end{equation}
The terms related to other sites can be obtained in a similar way. Therefore, expanding all terms in \eqref{first_2} and neglecting terms of order $\mathcal{O}(\Delta^3)$ gives
\begin{equation}
    \label{first_3}
    \dfrac{M}{6}\left[6CG(C)+\left(CG'(C)+\frac{G(C)}{1-C}\right)\sum_{k=1}^{6}\widetilde{C}_k+\left(\frac{G(C)}{(1-C)^2}+\dfrac{G'(C)}{1-C}+\frac{CG''(C)}{2}\right)\sum_{k=1}^6{\widetilde{C}_k^2}-6G(C)\mathcal{A}\right].
\end{equation}
Furthermore, since we have
\begin{equation}
\label{sum1}
\begin{aligned}
    \sum_{k=1}^{6}\widetilde{C}_k&=12\left(\frac{\partial^2C}{\partial x^2}+\frac{\partial^2C}{\partial y^2}\right)\frac{\Delta^2}{4}+\mathcal{O}(\Delta^3),\\
    &=12\mathcal{A}+\mathcal{O}(\Delta^3),
\end{aligned}
\end{equation}
and
\begin{equation}
\label{sum2}
\begin{aligned}
    \sum_{k=1}^{6}{\widetilde{C}_k}^2&=12\left(\left(\frac{\partial C}{\partial x}\right)^2+\left(\frac{\partial C}{\partial y}\right)^2\right)\frac{\Delta^2}{4}+\mathcal{O}(\Delta^3),\\
    &=12\mathcal{B}+\mathcal{O}(\Delta^3),
\end{aligned}
\end{equation}
Equation \eqref{first_3} becomes
\begin{equation}
    \label{become1}
    MCG(C)+M\left(2CG'(C)-G(C)+\frac{2G(C)}{1-C}\right)\mathcal{A}+M\left({CG''(C)}+\frac{2G(C)}{(1-C)^2}+\dfrac{2G'(C)}{1-C}\right)\mathcal{B}+\mathcal{O}(\Delta^3). 
\end{equation}
Remind that the second term in \eqref{delta1_s} is
\begin{equation}
    \label{become2}
    \begin{aligned}
    MCG(\newbar{K}_{\mathbf{s}}^{(\textls{m})})&=MCG(C)+MCG'(C)\widetilde{C},
    \\
    &=MCG(C)+MCG'(C)\mathcal{A}+\mathcal{O}(\Delta^3).
    \end{aligned}
\end{equation}
Then combining \eqref{become1} and \eqref{become2} gives
\begin{equation}
    \label{become3}
    \begin{aligned}
        \delta(C_{\mathbf{s}})&=\left(CG'(C)-G(C)+\frac{2G(C)}{1-C}\right)M\mathcal{A}
    +\left({CG''(C)}+\frac{2G(C)}{(1-C)^2}+\dfrac{2G'(C)}{1-C}\right)M\mathcal{B}+\mathcal{O}(\Delta^3),\\
    &=\left(CG'(C)+\frac{1+C}{1-C}G(C)\right)M\mathcal{A}
    +\left({CG''(C)}+\frac{2G(C)}{(1-C)^2}+\dfrac{2G'(C)}{1-C}\right)M\mathcal{B}+\mathcal{O}(\Delta^3).
    \end{aligned}
\end{equation}
Dividing both sides of the resulting expression by $\tau$, and letting $\Delta\to0$ and $\tau\to0$ jointly, with the ratio $\Delta^2/\tau$ held constant, leads to the following nonlinear reaction-diffusion equation,
\begin{equation}
    \label{become4}
    \frac{\partial C}{\partial t}=D_0\nabla\cdot\left[\left(CG'(C)+\frac{1+C}{1-C}G(C)\right)\nabla C\right]+\lambda CF(C),  
\end{equation}
where
\begin{equation}
\label{limitparameter1_s}
    D_0=\frac{M}{4}\lim_{\Delta,\tau\to0}
    \frac{\Delta^2}{\tau},\quad \lambda=\lim_{\tau\to0}\frac{P}{\tau}.
\end{equation}
If we define
\begin{equation}
    D(C)=CG'(C)+\frac{1+C}{1-C}G(C),
\end{equation}
then the continuum limit is written as
\begin{equation}
    \label{PDE_general_2D_sup}
    \frac{\partial C}{\partial t}=D_0\nabla\cdot\left[D(C)\nabla C\right]+\lambda CF(C).
\end{equation}
\newpage
\section{Numerical methods}
Here, we introduce the method of lines to numerically calculate solutions of the PDE
\begin{equation}
    \label{PDE_general_2D_s}
    \frac{\partial C}{\partial t}=D_0\nabla^2C+R(C),
\end{equation}
on a square domain $\Omega=\{(x,y),0<x<L,\ 0<y<L\}$.
We first discretise the spatial derivative in Equation \eqref{PDE_general_2D_s} with an $(I+1)\times (I+1)$ mesh. Nodes on the mesh are uniformly distributed with spacing $\delta x>0$ and indexed by $x_i$ and $y_j$ with $i=0,1,2,...,I$ and $j=0,1,2,...,I$ satisfying $I=L/\delta x$. We leave the time derivative continuous and obtain
\begin{equation}
    \label{methodofline}
    \begin{aligned}
    \frac{\textls{d}C_{i,j}}{\textls{d}t}=&\dfrac{D_0}{\delta x^2}(C_{i+1,j}+C_{i-1,j}+C_{i,j+1}+C_{i,j-1}-4C_{i,j})+R(C_{i,j}).
    \end{aligned}
\end{equation}
This equation is valid for interior nodes,  and is modified on the boundary nodes to simulate periodic boundary conditions. This system of $I\times I$ coupled ordinary differential equations is then integrated through time using MATLABs function ode45 \citep{ode45ref}. Following similar steps, we can also calculate the numerical solution of the PDE 
\begin{equation}
    \label{PDE_general_2D_new}
    \frac{\partial C}{\partial t}=D_0\frac{\partial^2 C}{\partial x^2}+R(C).
\end{equation}

\newpage
\section{Robustness of the stochastic simulations}
In this section we investigate the robustness of the stochastic simulations which generate discrete data from $V$ identically-prepared realisations. The variance of the total population density is
\begin{equation}
    \sigma^2=\frac{1}{V}\sum_{i=1}^{V}\left\lVert\left<C^i(t)\right>-\left<C(t)\right>\right\rVert,
\end{equation}
where $\left<C^i(t)\right>$ denotes the total population density estimated from the discrete simulation in the $i$th realisation and $\left<C(t)\right>$ denotes the averaged total population density. In practice, we consider $t\in[0,10^4]$. As we calculate the total population density via averaging 40 times identically-prepared realisations in the main document, we also consider $V=40$ in this section. We calculate the variance with the well-mixed, vertical strip and square initial distributions in Figure~\ref{Robustness}. For the well-mixed initial distribution,  $C(0)$ varies with $B\in[0.1,0.6]$. Note that we use $\mathcal{C}(0)$ as the $x$-axis in Figure~\ref{Robustness} as it is only equivalent to $C(0)$ for the well-mixed initial distribution. For the vertical strip initial distribution, we hold $B=1$ and change $\mathcal{C}(0)$ by varying the width of the strip. For the square initial distribution, we hold $B=1$ and change $\mathcal{C}(0)$ by varying the width of the square. The variance is small when $P/M=0.001$ for all three initial distributions in Figure~\ref{Robustness}(a), which indicates that averaging the data from our discrete simulations leads to a robust estimate of the  average occupancy. When we consider $P/M=0.02$, the variance is also low for vertical strip and square initial distributions. However, the variance becomes much larger when $\mathcal{C}(0)$ approaches $0.4$, which is the Allee threshold, in the well-mixed initial distribution. This is not surprising as the discrete simulations either lead to survival or extinction depending on fluctuations in the density.

\renewcommand{\thefigure}{S1}
\begin{figure}[t]
\centering
\includegraphics[width=0.9\textwidth]{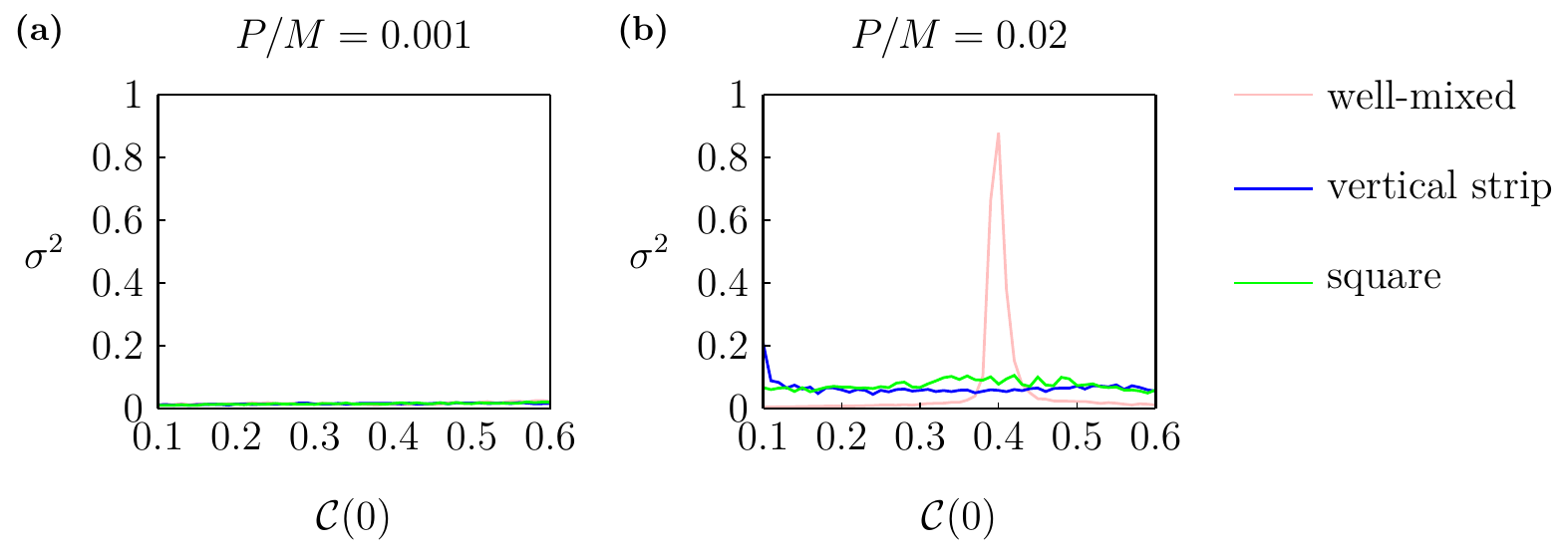}
\singlespace\caption{
\textbf{The robustness of the stochastic simulations.} (a) Variance of the difference between the averaged total population density, $\left<C(t)\right>$ and the total population density of the $i$th realisation, $\left<C^i(t)\right>$, where $i=1,2,...,40$, with $P/M=0.001$. (b) Variance of the difference between the averaged total population density, $\left<C(t)\right>$ and the total population density of the $i$th realisation, $\left<C^i(t)\right>$, where $i=1,2,...,40$, with $P/M=0.02$. Pink curves are generated with the well-mixed initial distribution. Blue curves are generated with the vertical strip initial distribution. Green curves are generated with the square initial distribution.} 
\label{Robustness} 
\end{figure}

\newpage
\section{Phase diagrams with \text{$B\ne 1$}}
Instead of varying the size of $\mathcal{H}$, we now vary $\mathcal{C}(0)$ by varying $B$. We constrain the region $\mathcal{H}$ as a vertical strip with width $w_1=64$, as shown in Figure~\ref{Phasediagram_3}(a). As $\mathcal{C}(0)=Bw_1/L$, where we fix $w_1/L=0.64$, the initial density $\mathcal{C}(0)$ varies from $0.192$ to $0.64$ when $B$ varies from $0.3$ to $1$ as illustrated in Figures~\ref{Phasediagram_3}(a)--(c). We vary $P/M=\lambda/(4D_0)$ by holding $M=1$ and varying $P\in[1/1000,4/100]$ and we discretise the $(B,P/M)$ space into a rectangular mesh with $36\times40$ nodes. Figure~\ref{Phasediagram_3}(d) shows a phase diagram illustrating how the survival probability, $S$, depends upon $B$ and $P/M$. The boundary that separates the eventual survival and extinction in the continuum model is shown in solid black, and the survival probability from the discrete simulations is shown in blue shading. The long–term predictions in terms of survival or extinction are consistent between the continuum and discrete models. Furthermore, we observe a different phenomenon compared to the results in Figure~10: There is a lower bound on $\mathcal{C}(0)$ for survival in Figure~\ref{Phasediagram_3}(d). This lower bound relates to $B=0.4$, and indicates the Allee threshold $A=0.4$. Unlike the solid line, $\mathcal{C}(0)=0.4$, which indicates a threshold of survival in the sense of global density, the dashed line, $B=0.4$, indicates a threshold of survival in the sense of local density.

\renewcommand{\thefigure}{S2}
\begin{figure}[t]
\centering
\includegraphics[width=0.75\textwidth]{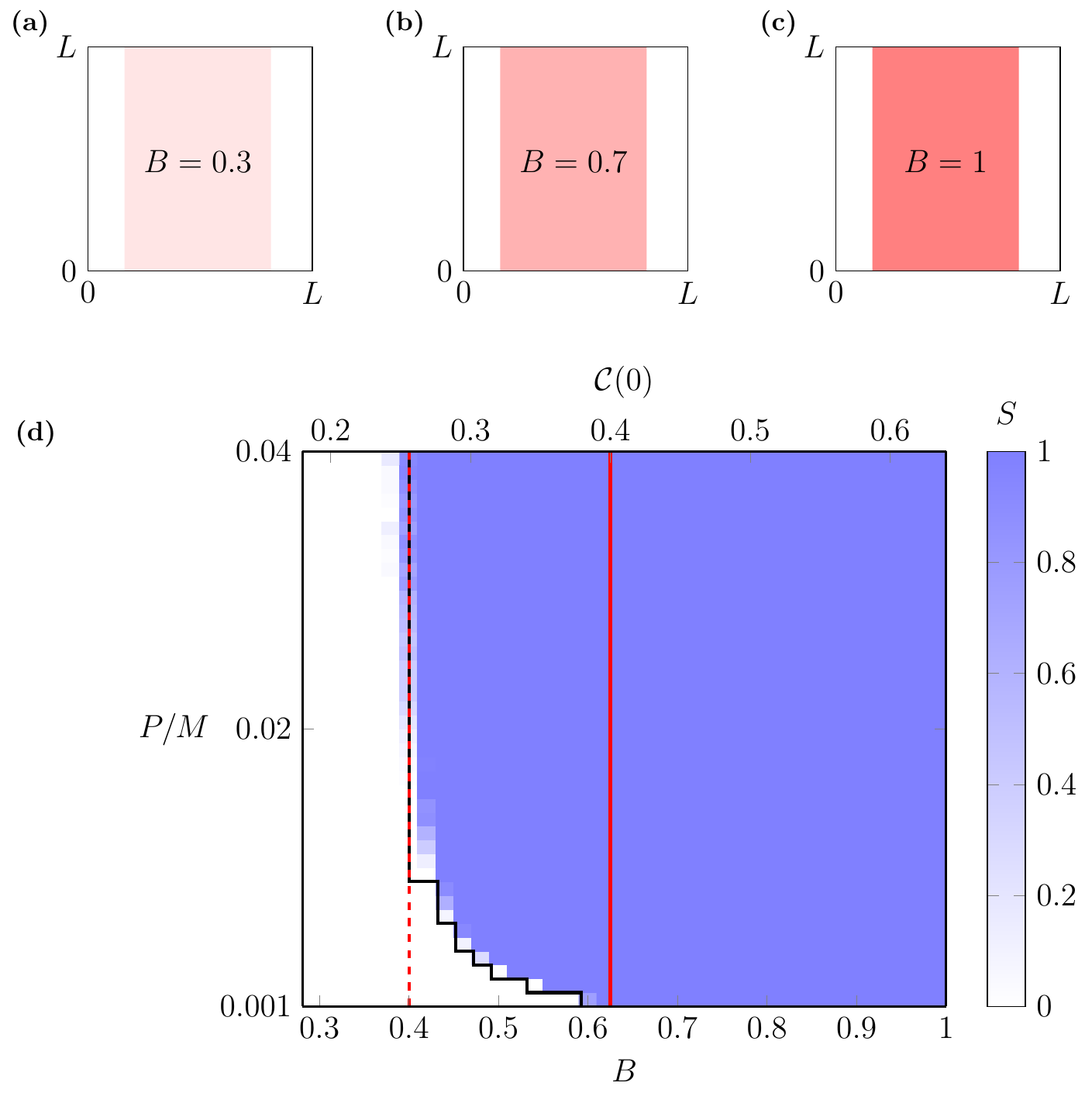}
\singlespace\caption{
\textbf{Phase diagram for survival/extinction with the vertical strip initial distribution}. (a)--(c) Three different initial distributions where $\mathcal{C}(0)=Bw_1/L$, and we fix $w_1=64$ and vary $B$. (d) Phase diagram on a rectangular mesh with $36\times40$ nodes for $B\in[0.3,1]$, $\mathcal{C}(0) \in [0.192, 0.64]$ and $P/M \in [1/1000,4/100]$. The black curve indicates the survival/extinction threshold from the continuum model and the blue shading shows the survival probability $S$ from the discrete simulations measured by 40 identically-prepared realisations. The vertical solid red line is $\mathcal{C}(0)=0.4$. The vertical dashed line is $B=0.4$. They both relate to the Allee threshold, $A=0.4$.}
\label{Phasediagram_3} 
\end{figure}

\newpage
Next, we consider the square initial distribution and constrain the region with width $w_1=80$ in Figures~\ref{Phasediagram_4}(a)--(c).  As $\mathcal{C}(0)=Bw_1^2/L^2$, where we fix $w_1^2/L^2=0.64$, the initial density $\mathcal{C}(0)$ varies from $0.192$ to $0.64$ when $B$ varies from $0.3$ to $1$. We again change $P/M=\lambda/(4D_0)$ by holding $M=1$ and varying $P\in[1/1000,4/100]$ and we discretise the $(B,P/M)$ space into a rectangular mesh with $36\times40$ nodes. With this initial condition we construct a phase diagram summarising the long–term survival outcomes as a function of $B$ and $P/M$ in Figure~\ref{Phasediagram_4}(d), which is very similar to the phase diagram in Figure~\ref{Phasediagram_3}(d) where we see that the long-term survival depends on $P/M$ and the two red lines indicating the Allee threshold.

\renewcommand{\thefigure}{S3}
\begin{figure}[t]
\centering
\includegraphics[width=0.8\textwidth]{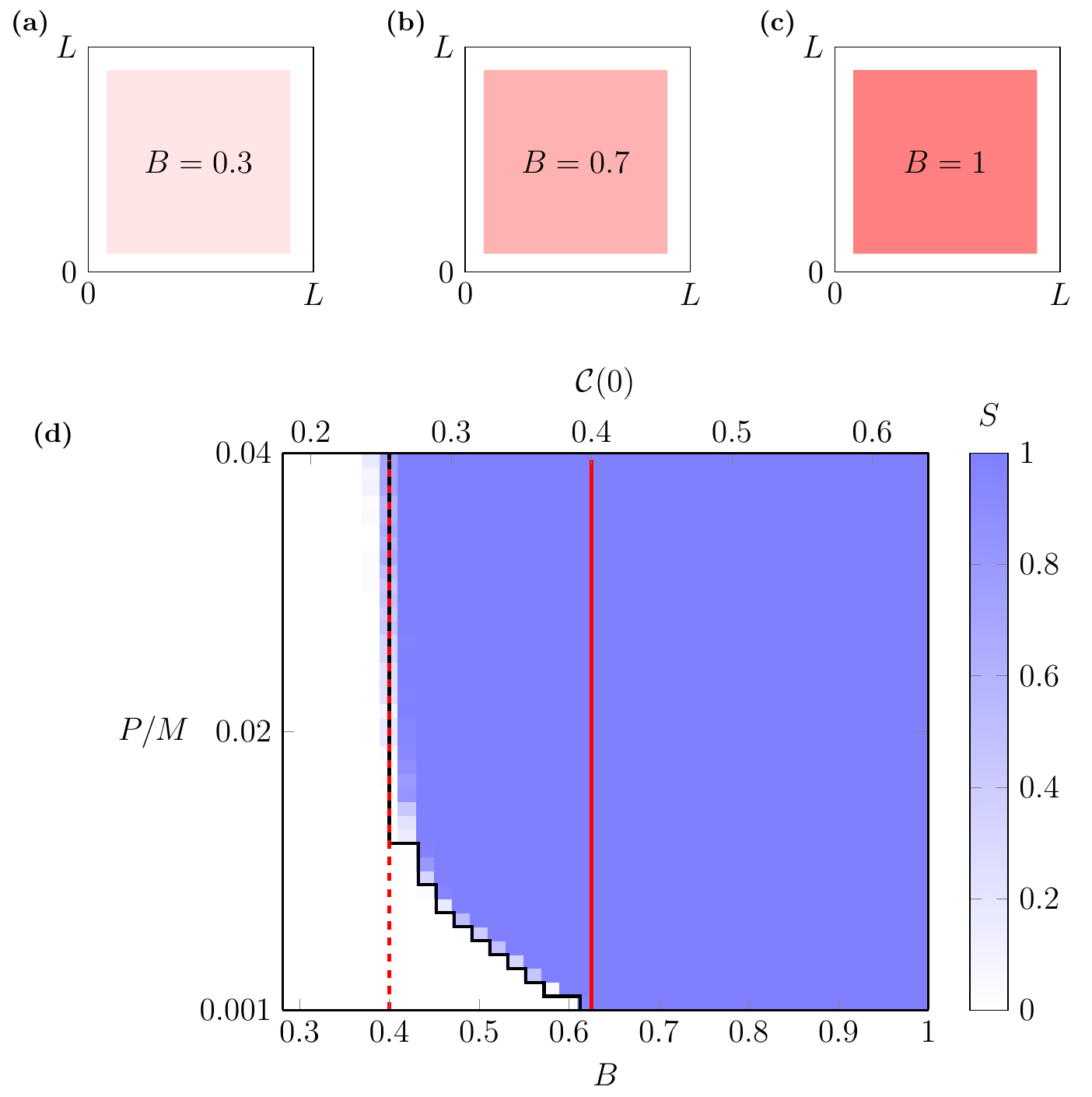}
\singlespace\caption{
\textbf{Phase diagram for survival/extinction with the square initial distribution}. (a)--(c) Three different initial distributions where $\mathcal{C}(0)=Bw^2/L^2$, and we fix $w=80$ and vary $B$. (d) Phase diagram on a rectangular mesh with $36\times40$ nodes for $B\in[0.3,1]$, $\mathcal{C}(0) \in [0.192, 0.64]$ and $P/M \in [1/1000,4/100]$. The black curve indicates the survival/extinction threshold from the continuum model and the blue shading shows the survival probability $S$ from the discrete simulations measured by 40 identically-prepared realisations. The vertical solid red line is $\mathcal{C}(0)=0.4$. The vertical dashed line is $B=0.4$. They both relate to the Allee threshold, $A=0.4$.}
\label{Phasediagram_4} 
\end{figure}

\newpage
To highlight the different fates of various populations, we compare the outcomes from the continuum model in Figure~\ref{012DCompare_1}(a), where we superimpose the boundaries that separate regions of survival and extinction for the well-mixed initial distribution (red), the vertical strip initial distribution (black) and the square initial distribution (green) described by Figure~1. Superimposing these curves divides the $(\mathcal{C}(0),P/M)$ plane into four regions with different long-term outcomes depending on the shape of the initial distributions. To emphasise these differences we compare solutions of the continuum model with different values of $\mathcal{C}(0)$ and $P/M$ in Figures~\ref{012DCompare_1}(b)--(g). The solutions in Figures~\ref{012DCompare_1}(b)--(g) correspond to various illustrative choices of $\mathcal{C}(0)$ and $P/M$. For example, the profiles in Figure~\ref{012DCompare_1}(b) related to region $\mathcal{R}_0$ all lead to extinction regardless of the shape of the initial distributions, whereas the profiles in  Figure~\ref{012DCompare_1}(c) related to region $\mathcal{R}_1$ lead to extinction for the well-mixed and square distributions, whereas the vertical distribution leads to survival.

\renewcommand{\thefigure}{S4}
\begin{figure}
\centering
\includegraphics[width=0.95\textwidth]{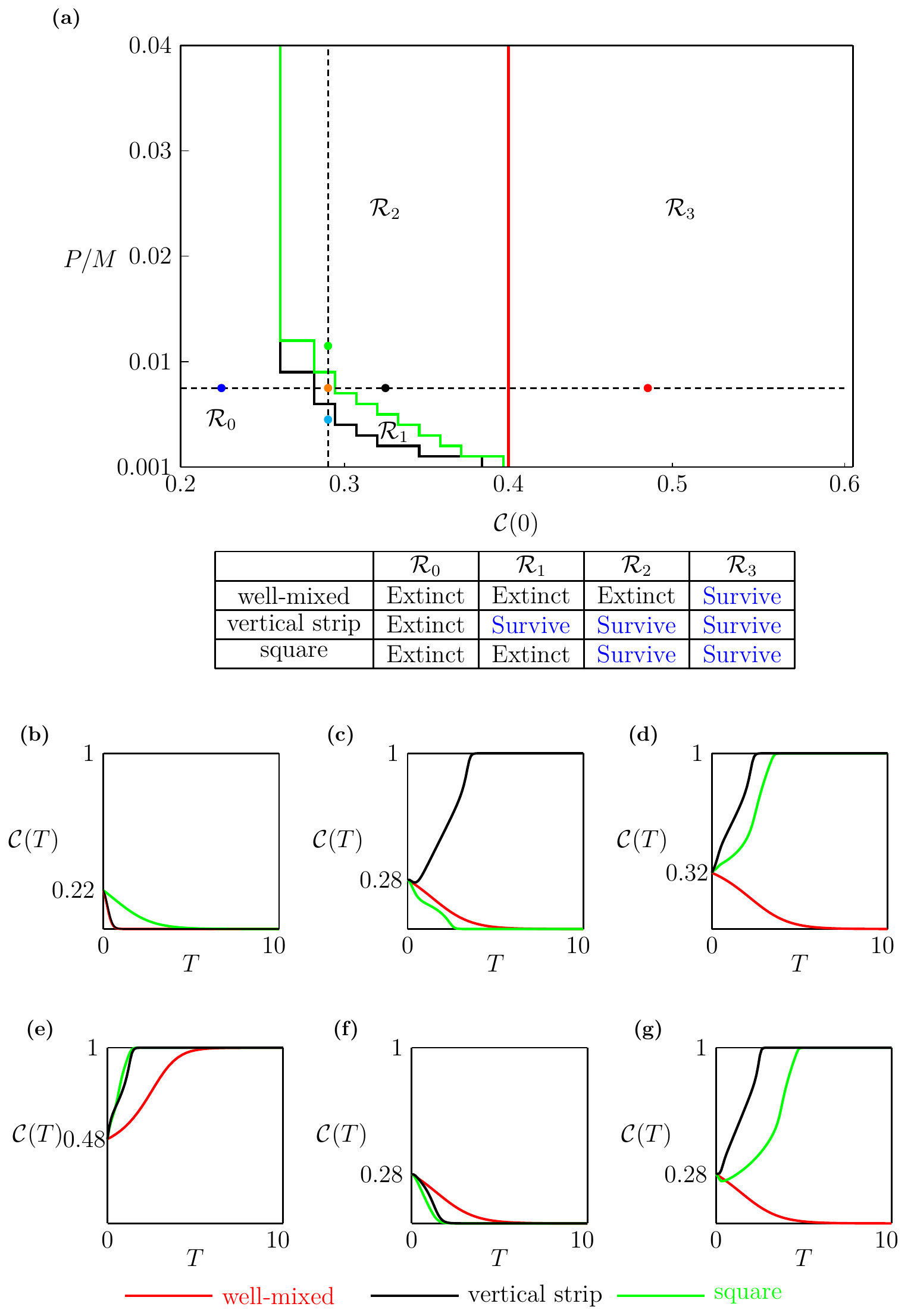}
\singlespace\caption{
\textbf{Role of dimensionality in long-term survival and extinction.} (a) The combined phase diagrams from the continuum model where the red, black and green curves highlight the boundaries separating extinction and survival for the well-mixed, vertical strip and square initial distributions, respectively.  (b)--(g) Profiles of $\mathcal{C}(T)$ for six different choices of $P/M$ and $\mathcal{C}(0)$. Parameters in (b)--(g) relate to the coloured discs superimposed in (a): (b) relates to the blue disc; (c) relates to the orange disc; (d) relates to the black disc; (e) relates to the red disc; (f) relates to the cyan disk and (g) relates to the green disc.}
\label{012DCompare_1} 
\end{figure}

\newpage

\section{Phase diagram with the no-flux boundary conditions}
We generate the survival/extinction threshold of the square initial distributions with no-flux boundary conditions along all boundaries, and compare it to the result obtained from the periodic boundary conditions in Figure~\ref{Phasediagram_noflux}. The survival/extinction threshold from the continuum model with the no-flux boundary conditions is the same as the result obtained from the periodic boundary conditions.

\renewcommand{\thefigure}{S5}
\begin{figure}
\centering
\includegraphics[width=0.8\textwidth]{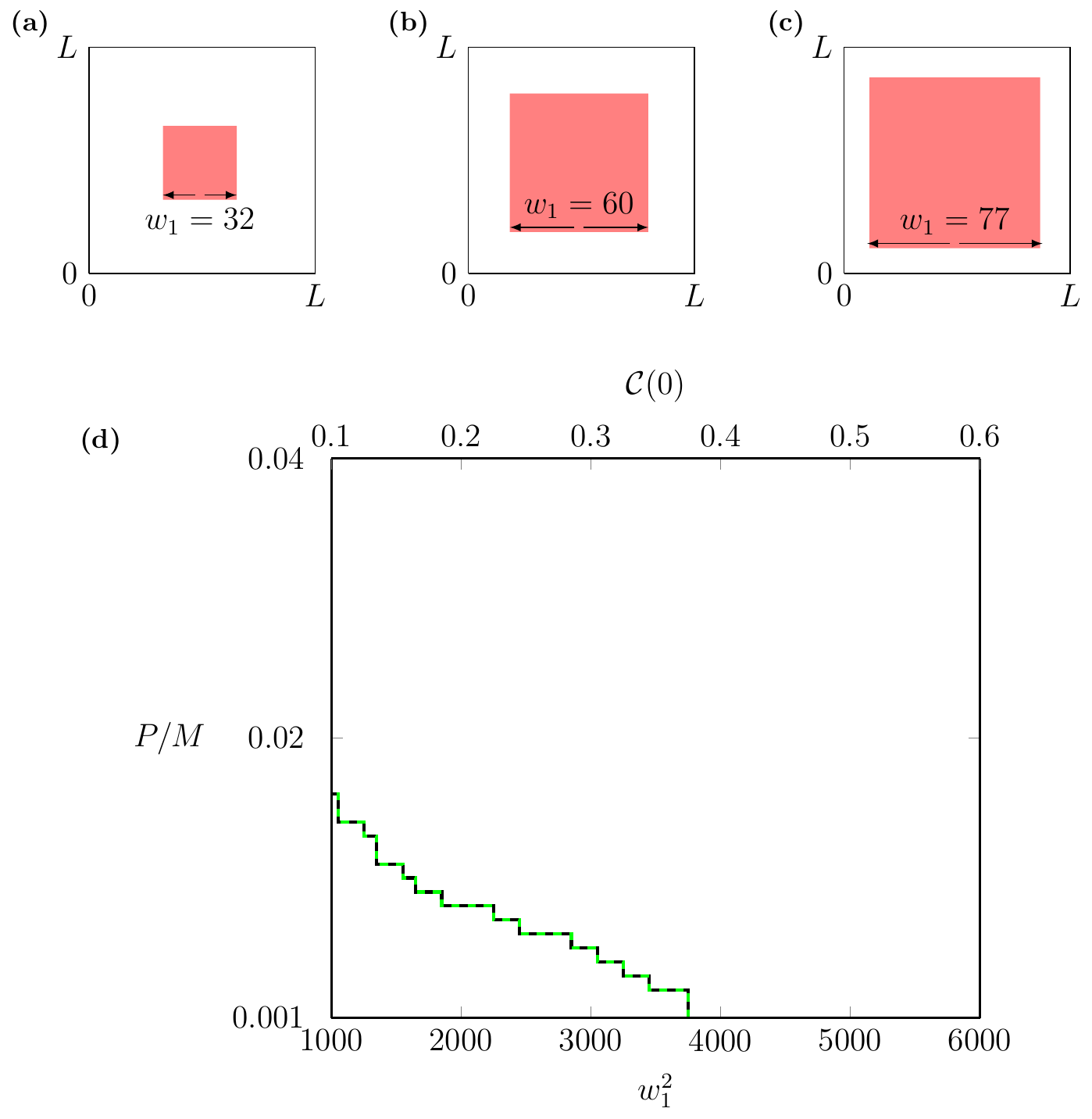}
\singlespace\caption{\textbf{Phase diagram for survival/extinction with the square initial distribution and different boundary conditions}. (a)--(c) Three different initial distributions where $\mathcal{C}(0)=w_1^2/L^2$, and we vary $w_1$. (d) Phase diagram of a rectangular mesh with $51\times40$ nodes for $w_1^2\in[1000,6000]$, $\mathcal{C}(0) \in [1/10, 6/10]$ and $P/M \in [1/1000,4/100]$ where $M=1$. The black solid curve is the survival/extinction threshold from the continuum model with the periodic boundary condition. The green dashed curve is the survival/extinction threshold from the continuum model with the no-flux boundary condition.}
\label{Phasediagram_noflux} 
\end{figure}

\newpage 

\section{Phase diagram with a larger domain}
We obtain the survival/extinction threshold from the square initial distributions on the domain with $L=200$, and compare it to the result obtained with $L=100$ in Figure~\ref{Phasediagram_largerdomain}. The boundaries separating survival and extinction with $L=100$ and $L=200$ are close to each other, especially when the size of the initially occupied region is small relative to the size of the whole domain. Furthermore, there are more cases of survival with $L=100$ when $P/M$ is small as the population spreads out across the domain rapidly, which indicates that the population is more susceptible to extinction in a larger domain.

\renewcommand{\thefigure}{S6}
\begin{figure}
\centering
\includegraphics[width=\textwidth]{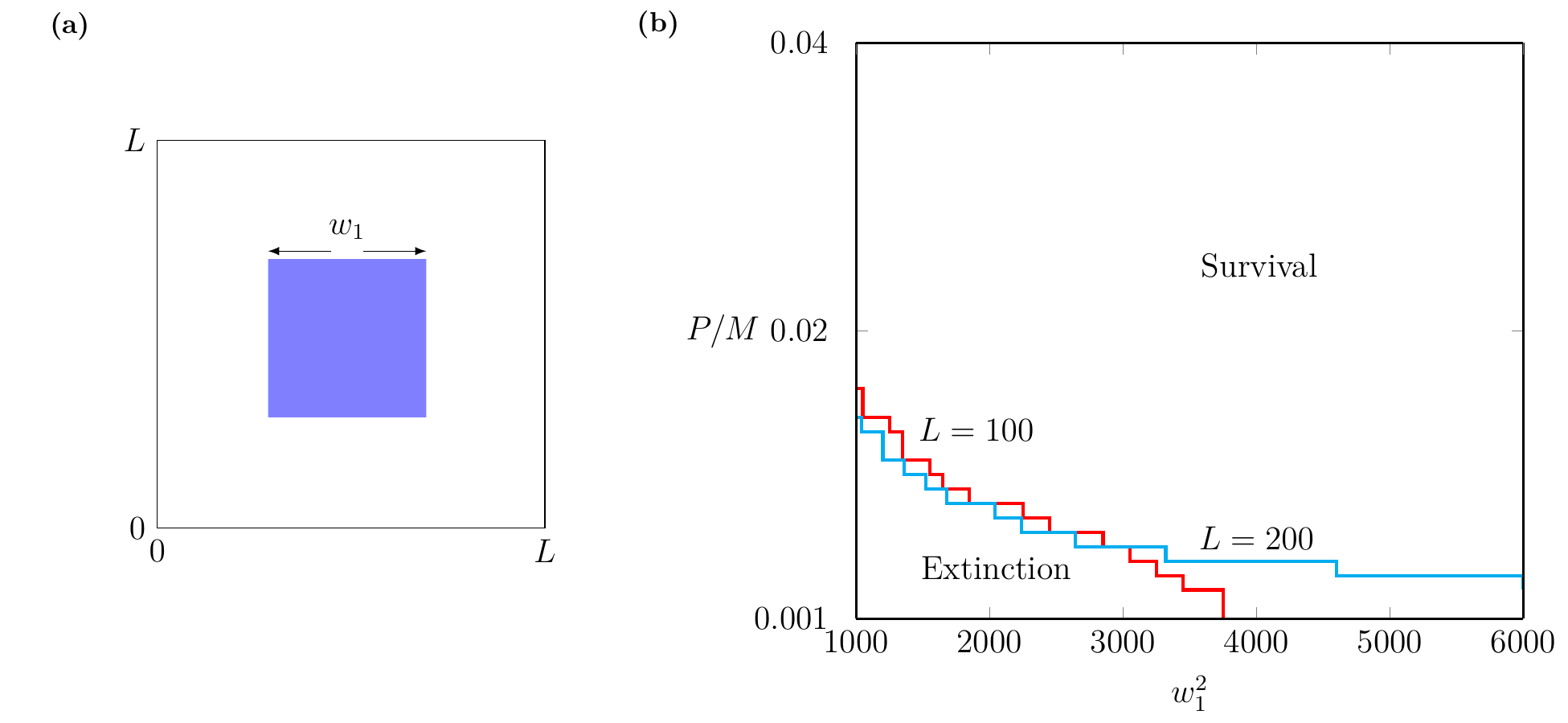}
\singlespace\caption{\textbf{Phase diagram for survival/extinction with the square initial distributions on different-sized domains}. (a) Square initial distribution with width $w_1$ on the $L\times L$ domain. (b) Phase diagram for $\mathcal{C}(0) \in [1/10, 6/10]$, $w_1^2\in[1000,6000]$ and $P/M \in [1/1000,4/100]$ where $M=1$. We vary the square initial distributions by varying $w_1$ on the domain with $L=100$ and $L=200$. The red curve is the survival/extinction threshold from the continuum model on the domain $L=100$. The cyan curve is the survival/extinction threshold from the continuum model on the domain $L=200$.}
\label{Phasediagram_largerdomain} 
\end{figure}

\newpage
\section{Value of the combined discrete-continuum framework}
Although the discrete simulations and continuous solutions match well in our framework, there is additional information in the discrete model that cannot be easily extracted from the continuum modelling approach. For example, it is unclear from the continuum model whether a population
survives mostly because of a high number of births, or because of a low number of deaths. Tracking the age of
agents in the discrete model gives indications as to where agents of different ages are located in space. In Figures~\ref{fig:14}(a)--(d), we show the age group of a population in one realisation with vertical strip and square initial distributions. Both initial distributions lead to the survival of populations, but the age structure of the two populations is different, see Figure~\ref{fig:14}(e) for the number of agents in different generations when the population occupies the whole domain. 

\renewcommand{\thefigure}{S7}
\begin{figure}[]
\centering
\includegraphics[width=\textwidth]{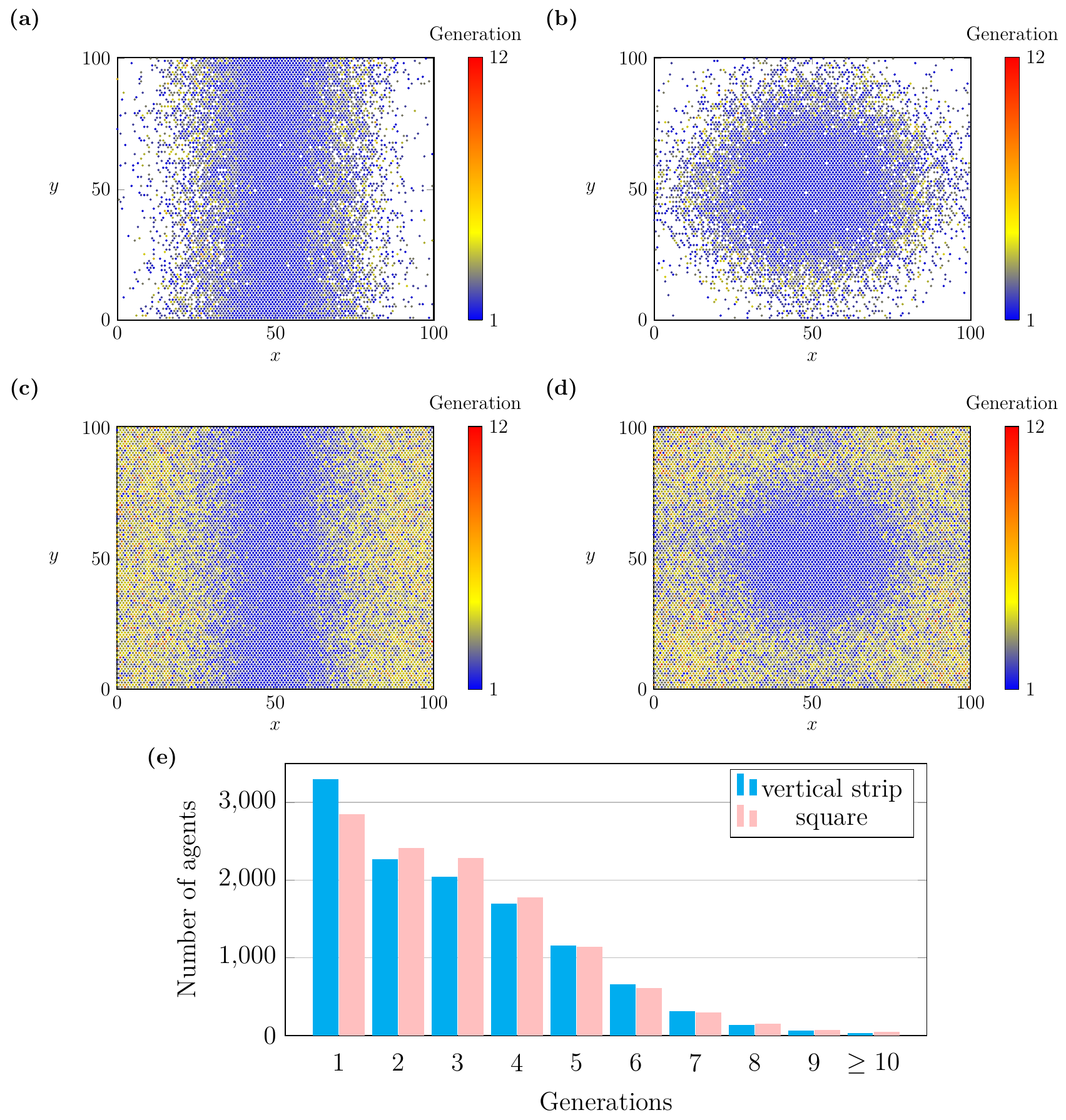}
\singlespace\caption{\textbf{The age group of the population with vertical strip and square initial distributions where $\mathcal{C}(0)=0.4$ and $P/M=0.01$, where $P=0.01$ and $M=1$}. (a) Snapshot of the discrete simulation with the vertical strip initial distribution at $t=1000$. The agents initially placed on the domain are the first generation (blue). (b) Snapshot of the discrete simulation with the square initial distribution at $t=1000$. (c) Snapshot of the discrete simulation with the vertical strip initial distribution when the population occupies the whole domain. (d) Snapshot of the discrete simulation with the square initial distribution when the population occupies the whole domain. (e) The number of agents in different generations.}
\label{fig:14} 
\end{figure}

Another advantage of the discrete model is the ability to track the individual behaviours of the agents. For example, in the discrete model one can trace the trajectory of individuals, which provides insights into the motility mechanisms \citep{Cai2006motility,simpson2009pathlines}. In Figure~\ref{fig:15}, we initially select five agents and trace them until the population occupies the whole domain or until the agent dies. We observe that those agents closer to free space are more motile. Furthermore, once they move into a less crowded region, they have higher probability of dying. This phenomenon is consistent with the growth mechanism governed by the strong Allee kinetics. 

\renewcommand{\thefigure}{S8}
\begin{figure}[]
\centering
\includegraphics[width=0.8\textwidth]{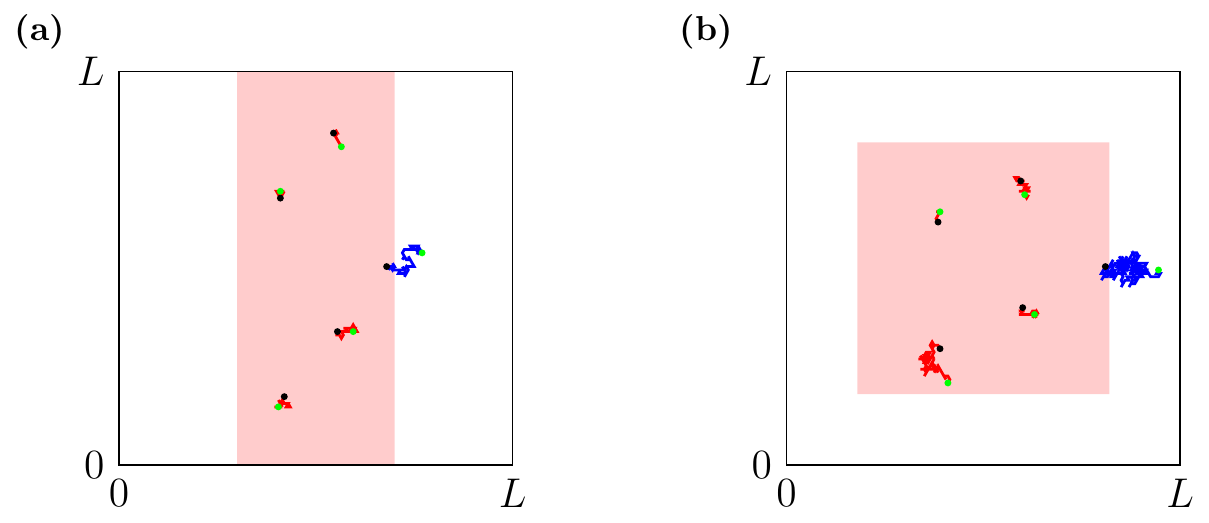}
\singlespace\caption{\textbf{Trajectories of agents in discrete simulations with the vertical strip and square initial distributions where $\mathcal{C}(0)=0.4$ and $P/M=0.01$, where $P=0.01$ and $M=1$}. (a) Trajectories of five agents with the vertical strip initial distribution. (b) Trajectories of five agents with the square initial distribution. The pink area indicates the initially occupied region with $B=1$. The black points indicate the initial positions of the five individuals. The green points indicate the final positions of the five individuals when the population occupies the whole domain. Note that both agents whose trajectories are highlighted in blue in (a) and (b) died before the population occupies the whole domain .}
\label{fig:15} 
\end{figure}
\newpage
%\todo{CAn you please go over the reference list carefully and please make sure the url are not double and all working! one url per paper and one DOI; if published arxiv link should not be there anymore}
\bibliographystyle{elsarticle-harv.bst} 
\bibliography{Paper3_arXiv.bib}

\end{document}